\documentclass[preprint,12pt]{elsarticle}

\usepackage{amssymb}
\usepackage[linesnumbered,ruled,vlined]{algorithm2e}%

\usepackage{url}
\usepackage[usenames]{color}
\definecolor{linkcolor}{cmyk}{0.98,0.13,0,0.43} % midnight blue

\graphicspath{{figures/}}
\DeclareGraphicsExtensions{.pdf}
\usepackage{comment}

\usepackage{lipsum} % For Tables
\usepackage{tabularx}
\usepackage{lipsum,amsmath,multicol}
\usepackage{multirow}
\usepackage{array,booktabs}
\usepackage{lipsum} % For Tables
\usepackage{tabularx}
\usepackage{adjustbox}
\usepackage{lipsum,amsmath,multicol}
\usepackage{multirow}
\usepackage{array,booktabs}
\usepackage{threeparttable} 
\usepackage{array}
\usepackage{makecell}

\usepackage{pbox}

\usepackage{csquotes}

\usepackage[caption=false,font=footnotesize]{subfig}
\usepackage{float}
\newboolean{printComments}
\setboolean{printComments}{true}

\newcommand{\yasr}[1] {{\ifthenelse {\boolean{printComments}} {\color{red}#1}{}}}
\newcommand{\yasb}[1] {{\ifthenelse {\boolean{printComments}} {\color{blue}#1}{}}}

\usepackage{geometry}
\journal{Journal of Network and Computer Applications}

\begin{document}

\begin{frontmatter}



% Adel Nadjaran Toosi 

\title{An Automated Implementation of Hybrid Cloud for Performance Evaluation of Distributed Databases}

\author{Yaser Mansouri, Victor Prokhorenko, and M. Ali Babar}

\address{Centre for Research on Engineering Software (CREST) Lab.}
\address{School of Computer Science}
\address{The University of Adelaide, Adelaide, Australia}

\begin{abstract}
A Hybrid cloud is an integration of resources between private and public clouds. It enables users to horizontally scale their on-premises infrastructure up to public clouds in order to improve performance and cut up-front investment cost. This model of applications deployment is called cloud bursting that allows data-intensive applications especially distributed database systems to have the benefit of both private and public clouds. 

In this work, we present an automated implementation of a hybrid cloud using (i) a robust and zero-cost Linux-based VPN to make a secure connection between private and public clouds, and (ii) Terraform as a software tool to deploy infrastructure resources  based on the requirements of hybrid cloud. We also explore performance evaluation of cloud bursting for six modern and distributed database systems on the hybrid cloud spanning over local OpenStack and Microsoft Azure. Our results reveal that MongoDB and MySQL Cluster work efficient in terms of throughput and operations latency  if they burst into a public cloud to supply their resources. In contrast, the performance of Cassandra, Riak, Redis, and Couchdb  reduces if they significantly leverage their required resources via cloud bursting.  
\end{abstract}


\end{frontmatter}


\section{Introduction}\label{sec:introduction}
Cloud computing is the mainstream delivery of on-demand and easy-to-use services such as computing, networking, storage, databases,  and software over the Internet \cite{buyya2009}. It provides a pay-per-usage model in which consumers typically pay for cloud services they use. Cloud computing comes traditionally into two models \cite{armbrust2010}: \textit{public} and  \textit{private}. A public cloud provides computing, storage and networking resources to the general public over Internet while a private cloud facilitates resources  of on-premises infrastructure for the dedicated use of a specific organization.  

A hybrid cloud is a seamless integration of public and private clouds to take the best of both worlds \cite{rimal2009}. It enables \textit{cloud bursting} in which applications initially leverage private cloud and burst into a public cloud when private resources are not enough to provision under spiking  workload. A hybrid cloud significantly benefits its owner in terms of security in compliance with the location of sensitive data, availability, reliability and cost reduction. However, \textit{construction of hybrid clouds} and \textit{deployment of applications on top of them} are not trivial tasks. We investigate both tasks in this work. 

\textit{In respect to the hybrid cloud implementation, the first issue we intend to address is a secure, robust and cost-effective connection between public and private clouds.} Construction of hybrid clouds  demands a secure connection between computational resources scattered across private and public  clouds under different administrative domains.  It means that different virtual machines/nodes in sub-networks need to be  connected. This connectivity implies that two separate sets of IP ranges should be connected to make an automated resource provisioning across private and public clouds. One viable solution is to allocate public IPs to VMs to  make connections between clouds over Internet. Provided public IPs in a specific range for a private cloud is infeasible in most cases. Moreover, hybrid clouds require a secure connection channel especially for mission-critical systems in which secure data is transmitted over Internet.

Network virtualization techniques allow a secure connection channel between private and public clouds through building an overlay network over the Internet. Virtual Private Networks (VPNs) provide such overlay network to make a secure connection between VMs in the hybrid cloud. The well-known cloud providers offer different VPN connectivity options to make a secure connection across clouds [5].  Such VPN options may not be compatible, robust (observed for Azure VPN), and suffer from high monetary cost. In contrast, WireGuard\footnote{WireGuard: \url{https://www.wireguard.com/}} was initially introduced in 2016 as a Linux kernel-based VPN that provides a \textit{flexible}, \textit{robust}, \textit{interoperable}, \textit{secure} and \textit{zero-cost connection} between private and public clouds according to individual needs.

\textit{In regard to the hybrid cloud implementation, the second issue we want to figure out is to deploy infrastructure and applications in an automated manner.} Automation of a hybrid cloud implementation is the key concept to deploy infrastructure resources, to make a secure connection between clouds, and to route data among sub-networks across clouds. Such automation allows users to quickly exploit resources with different flavors, reduces  human's work, and consistently creates an environment to deploy the desired big data applications. We  leveraged Terraform\footnote{Terraform: \url{https://www.terraform.io/}} as a software tool to provide an automated implementation of the hybrid cloud based on the required VMs size, VMs number, sub-network features (i.e., ingress and egress port number) and so forth.

Today's IT systems may benefit from hybrid clouds to boost performance, manage business velocity,  avoid data lock-in, and keep sensitive data on-premise and insensitive data in  public clouds. The performance of such systems depend on the cloud-based distributed databases that might be different in serving read and write operations, and  managing data replication, data sharding, and data consistency. These databases are classified into relational and NoSQL\cite{Mansouri2017a}. Relational databases have pre-defined scheme and possess a mature market. By contrast, NoSQL databases have a schema-less data model and compromise strong consistency for better performance. 

%Due to ever-increasing  demands in data volume, variety, and velocity, so-called 3 Vs in big data, relational databases are hardly applicable to such data. To fulfil such requirements, we need to deploy distributed, schema-less, and scalable databases to store and manage big data. 

Due to ever-increasing demands in data Volume, Variety, and Velocity (i.e., 3 Vs in big data), relational databases might not be applicable to such data and NoSQL databases seem a viable solution \cite{Han2011}. Currently, there are more than 225 NoSQL databases\footnote{NoSQL databases: \url{http://nosql-database.org/}} widely used in large-scale companies like Facebook, Twitter, Amazon, and Google. These companies provide infrastructure for small- and medium-size businesses to run NoSQL databases. Nevertheless, such businesses do not completely store their data on public infrastructure (provided by these companies) and exploit their private infrastructure to achieve their needs in performance and security aspects. Due to increasing deployment of NoSQL databases in different industry domains like e-commerce, IoT, blockchain, social networks, to name a few, the evaluation of cloud bursting at the level of NoSQL databases not only attracts research attention but is also useful from industry perspective to investigate the impact of cloud bursting on distributed databases.

Integrating private and public clouds to build a hybrid cloud was initially conducted in the academic domain  to evaluate the performance of different applications \cite{NADJARANTOOSI2018}\cite{Calheiros2011}. Due to potential benefits of hybrid clouds, the well-known cloud providers offer developers rich platforms to run applications in an on-premises environment and deliver the same services as in the public clouds. In fact, such platforms make easy to build up a hybrid cloud. Microsoft Azure released Azure Stack Development Kit (ASDK)\footnote{ASDK:\url{https://azure.microsoft.com/en-us/overview/azure-stack/}} in 2017, following \textit{Outposts}\footnote{Outposts: \url{https://aws.amazon.com/outposts/}} and \textit{Anthos}\footnote{Anthos: \url{https://cloud.google.com/anthos/}} by AWS and Google in 2018. Neither commercial nor academic projects reveal how they support hybrid clouds. 
None of these projects have  revealed how to build up a general architecture for hybrid clouds. Also, due to the nature of commercial products, cloud providers only offer these products as black-boxes  and developers can deploy them as a hybrid cloud in  conjunction with public clouds.  %In academic domain, authors developed a specific hybrid cloud based on their needs and they have not revealed how to build hybrid cloud. Recently, Nadjaran Toosi et al. [???] build up a hybrid cloud setup between Microsoft Azure and Aneka worker [???]  to evaluate  resource provisioning and scheduling algorithm. In industry domain, 

There is a large body of literature on deployment of data-intensive applications on hybrid clouds \cite{NADJARANTOOSI2018}\cite{CALHEIROS2012}\cite{VECCHIOLA2012}\cite{Zhou2019}. These studies focused on reduction of execution time for data-intensive applications within constrained time, monetary cost, or both. Several of  these studies measured performance evaluation in hybrid clouds. However, they neither leveraged a cost-free, secure, and resilient VPN to build a hybrid cloud, nor evaluated distributed databases on the hybrid clouds. By contrast, several studies evaluated the performance of distributed databases either on the public clouds \cite{Kuhlenkamp2014}\cite{Klein2015} or private clouds \cite{Abramova2013}\cite{Lima2016}\cite{Li2013}. None of these studies have specifically assessed the performance impact of cloud bursting on distributed databases running in hybrid clouds to comprehend how well these databases work on this model of clouds. In contrast to the public and private clouds offering all resources at the same datacenter, we strongly believe that distance between the private and public cloud datacenters has significant effects on the performance of distributed databases. Therefore, all these gaps discussed above motivate us to make the fallowing contributions.

%and evaluation of distributed databases on either public clouds \cite{Rabl2012} or private clouds \cite{Abramova2013}\cite{Lima2016}\cite{Li2013}. None of these studies have specifically assessed the performance impact of cloud bursting on distributed databases in hybrid clouds to comprehend how well distributed databases work on such infrastructure.   
%To fill the gaps delineated above, we make the following contributions:
\begin{itemize}
    \item We present different usage models of hybrid clouds, and for each model we identify purpose, key challenges, optimization domain, features and applicability.
    \item We automate the implementation of the hybrid cloud using Terraform\footnote{Terraform: \url{https://www.terraform.io/}} as an automation tool to provision and manage the infrastructure of clouds. In our implementation, we use WireGuard, as  Linux kernel-based VPN, to make a secure, robust, and zero-cost connection between public and private clouds. 
    \item We conduct experimental cloud-bursting evaluation at the database level and report on the details of our experience with this implementation from an industry view point.
\end{itemize}
The rest of the paper is organized as follows. Section 2 presents the related work. Section 3 delineates different usage models of hybrid clouds, following by an automated implementation of the hybrid cloud in Section 4. Section 5 describes general features of distributed databases and workload setup. Performance evaluation is discussed in Section 6. Finally, Section 7 concludes the paper and  outlines the future research directions.

\begin{table*}[t]
	
	\begin{threeparttable}
	\caption{Comparison of relevant studies with our work}\label{tab:relatedwork}
	\centering
	%\vspace{-3mm}
	\tiny
	\begin{tabular}{p{3.5cm} p{1cm}p{1.2cm}p{1.5cm}p{1cm}p{1cm}p{5.7cm}}
		%\centering
		\hline
		              Paper               &Application  &Cloud     &Network connection &Cloud bursting &Evaluation  &Objectives\\\hline\hline
		Toosi et al.\cite{NADJARANTOOSI2018}      &BoT  &Hybird    &Public VPN &Yes    &Imp. &Providing deadline requirements for data-intensive applications     \\
		Calheiros et al.\cite{CALHEIROS2012}      &BoT  &Hybird    &NS  &Yes    &Imp.  & Increasing the capacity of Desktop Grids via Cloud               \\
		Vecchiola et al.\cite{VECCHIOLA2012}      & BoT  &Hybird   &NS &Yes    &Imp. & Reduction in execution time via Ankeka and Cloud\\
		Tuli et al. \cite{Tuli2020}               &BoT  &Hybrid    &Public VPN   &Yes    &Imp. &Reduction in bandwidth consumption via considering sharing files\\\hline
		Li et al. \cite{Li2018}                   &OSN   &Hybrid   &IPsec tunnel  &Yes    &Imp. & Reduction in waiting and response times in a private cloud\\
		Moschakis and Karatza\cite{Moschakis2015} &BoT   &Hybrid    &NA   &Yes       &Sim.  &Optimization of performance and cost via scheduling tasks\\
		Abdi et al.\cite{Abdi2017}                &BoT    &Hybrid   &NA   &Yes      &Sim. & Minimizing the total cost of applications in cloud federation  \\
	    Zhou et al.\cite{Zhou2019}                &Workflow &Hybrid	&NA   &Yes     &Sim.  & Optimization monetary cost and makespan of workflow applications\\\hline
		Rabl et al.\cite{Rabl2012}                &DDB    &Private   &NA  &No      &Imp. & Performance evaluation of Cassandra, HBase, Redis, Voldemort, VoltDB, MySQL\\ 
		Kuhlenkamp et al.\cite{Kuhlenkamp2014}    &NSDB   &Public   &NA &No    &Imp. &Scalability and elasticity evaluation of Cassandra and HBase\\
		Li et al.\cite{Li2013}                    &NSDB   &Private  &NA &No    &Imp. & Evaluation of MongoDB, RavenDB, CouchDB, Cassandra, Hypertable, Couchbase, and MySQL in performance\\
		Klein et al.\cite{Klein2015}              &NSDB   &Public  &NA  &No    &Imp. & Evaluation of MongoDB, Cassandra, and Riak in consistency support\\
		Abramova and Bernardino\cite{Abramova2013} &NSDB  &Private  &NA &No &Imp. & A comparison between Cassandra and MongoDB in performance\\
		\textbf{This Work}                        &DDB    &Hybrid   &WireGuard &Yes & Imp. & Implementation of a hybrid cloud in the network level and evaluation of DDBs in the context of cloud bursting      
		\\\hline

	\end{tabular}
	\begin{tablenotes}
      \small
      \item  Abbreviations: BoT: Bag of Task; OSN: Online Social Network; DDB: Distributed Databases; NSDB: No-SQL Databases; NS: Not Specified;\\ Imp: Implementation; Sim: Simulation
   \end{tablenotes}
	\end{threeparttable}
	\vspace{-5mm}
\end{table*}

\section{Related Work}\label{sec:related}

\textbf{Hybrid cloud, benefit and challenges:} A Hybrid cloud integrates on-premises infrastructure (i.e., a private cloud) and a public cloud like Microsoft Azure and Amazon Web Services (AWS). It gives an organization an ability to scale its on-premises infrastructure up to handle workload spikes that demand more resources than available locally. A hybrid cloud brings the “best of both worlds” in the following features.

\textit{Security}: A Hybrid cloud enables organizations to split the data into  \textit{sensitive}  and \textit{insensitive} for storing  respectively in the private and public clouds. 	\textit{Resilience}: As the ability to guarantee an acceptable level of service in the case of faults and challenges to normal operations. Resilience of a  hybrid cloud depends on the infrastructure of clouds (guaranteed by cloud providers) and the network connection between clouds (guaranteed by VPN connection). To make a resilient and robust connection between clouds, we exploit WireGuard. \textit{Scalability and Reliability}: A hybrid cloud allows organizations to add/remove resources to/from their pool of resources as workload changes. It also offers higher reliability in the case of natural disasters as users can recover  data from a backup if data was replicated in both clouds. \textit{Cost saving}: A hybrid cloud mitigates capital expenditures to handle short-term spikes in workloads that require resources beyond the available ones in a private cloud. Nevertheless, a connection between private and public clouds can be costly. These features  can be achieved through a secure, resilient and cost-free VPN (e.g.,WireGuard) that connects the private and public clouds involving in the hybrid cloud.

Being no exception, a hybrid  cloud arises several challenges in two dimensions. The first one is implementation of hybrid clouds for which robust, automated, secure and cost-effective connection, network routing, and provisioning cloud infrastructure should be figured out.  The second dimension  is the deployment of applications in hybrid clouds for which automated resource discovery and  optimized Quality of Service(QoS)-- such as monetary cost and response time for running applications on a hybrid cloud-- should be addressed. In summary, these challenges depend on the usage models of hybrid clouds discussed later.
There is a plethora of literature that addresses these challenges in the context of the hybrid clouds. To position our work, we divide related work into the following subsections.

\textbf{Hybrid cloud implementation}:
Due to strong desire to exploit hybrid clouds by IT businesses, recently the well-known cloud providers enable deployment of their native services to on-premises infrastructure. This brings a potential capability to implement a truly consistent hybrid cloud through VPN. In this respect, Microsoft Azure offered Azure Stack Development  Kit (ASDK)  in  2017. Due  to cut-throat competition among cloud providers, AWS and Google respectively offered \textit{Outposts} and \textit{Anthos} in 2018. In spite of such benefits, these  products are not lightweight in hardware requirements and incur heavy costs through using VPNs to build hybrid cloud. 

 Before releasing these commercial products, researchers have conducted rich research on hybrid clouds. They evaluated their proposed algorithms, policy, and methods through simulation and implementation. Toosi et al. \cite{NADJARANTOOSI2018} recently set up a hybrid cloud including Microsoft Azure and two PC workers to evaluate their data-aware resource provisioning algorithm. Tuli et al. \cite{Tuli2020} designed dynamic resource provisioning and task scheduling algorithms and evaluated on the same setup of a hybrid cloud in \cite{NADJARANTOOSI2018} but with different VM flavors. \cite{CALHEIROS2012} and \cite{VECCHIOLA2012}  have conducted almost similar setup to evaluate an algorithm leveraging dynamic resources to meet the deadline requirements of Bag-of-Task. Li et al. \cite{Li2018} proposed  a cost-aware job scheduling approach based on queuing theory in hybrid clouds in a static setup using a public VPN. Differently, \cite{Loreto2015} implemented a software layer on top of hybrid cloud infrastructure to dynamically  deploy and scale virtual clusters.

\textbf{The performance impact of cloud bursting on distributed database systems}: Cloud bursting allows an application to run in a private cloud and burst into a public cloud in case the demand for computing and storage resources spikes. One suitable candidate for such deployment is distributed databases which are different in data model, data sharding, data replication and consistency. 

NoSQL databases tend to be a better solution for modern big data applications. They represented themselves as alternatives that can store big data characterized by 3 Vs: \textit{volume}, \textit{veracity} and \textit{variety}.  Due to a variety of NoSQL databases with different features, it is crucial to evaluate the performance impact of cloud bursting on such databases.

Initially in \cite{Li2013}, the performance of some distributed databases have been investigated for read, write, delete, and instantiate operations. Rabl et al. \cite{Rabl2012} evaluated  four NoSQL databases (Voldermort\footnote{Voldermort: \url{https://www.project-voldemort.com/voldemort/}}, HBase\footnote{HBase:\url{https://hbase.apache.org/}}, 
Cassandra\footnote{Cassandra: \url{http://cassandra.apache.org/}}, and 
Redis\footnote{Redis:\url{https://redis.io/}}, and two relational databases (MySQL cluster\footnote{MySQL:\url{https://www.mysql.com/}} and VoltDB\footnote{VoltDB:\url{https://www.voltdb.com/}}) with focus on performance and scalability. Kuhlenkamp et al. \cite{Kuhlenkamp2014} conducted experiments on NoSQL databases (HBase  and Cassandra) in terms of scalability and elasticity. In these studies, all experimental evaluations have been conducted on either a public or private cloud. Thus, they neither deal with the implementation of a hybrid cloud nor evaluate distributed databases in  this model of cloud. 

%Towards operational cost minimization for cloud bursting with deadline constraints in hybrid clouds
%Optimal provisioning and scheduling of analytics as a service in cloud computing
%Shared Data-Aware Dynamic Resource Provisioning and Task Scheduling for Data Intensive Applications on Hybrid Clouds using Aneka

Besides distributed databases, other type of data intensive applications have been recently evaluated on hybrid clouds. Xu \cite{xu2017} developed a cost and energy aware data placement method for privacy-aware big data applications. \cite{Moschakis2015}, \cite{Abdi2017} and \cite{XIONG2016}  investigated time- and cost-constrained algorithms to handle cloud-bursting using simulators like Cloud-Sim. Zhou et al. \cite{Zhou2019} proposed an approach that optimizes the monetary cost of scheduling workflows under constrained time, and then extended it to optimize both makespan and monetary cost for workflow scheduling. They achieved up to 100\% cost saving and an effective cost-makespan trade-off in comparison to the competing approaches through a simulation platform. In contrast to our work,  these studies all have investigated the effects of methods in a simulated testbed.

In comparison to the discussed studies in Table \ref{tab:relatedwork}, our work is different in two aspects.
(i) we have automated the implementation of hybrid cloud at the network level using Terraform as an automation tool and WireGuard as a secure, robust, and zero-cost VPN. This automated implementation of the hybrid cloud allows us to dynamically, consistently and repeatedly (with the least human interference) exploit infrastructure resources based on the desired features such as VMs number, VMs size, VMs region, Network features, database installation, database cluster configuration, and so on. (ii) We have evaluated the impact of cloud bursting on the performance of distributed databases.
%We summarize all the discussed studies in Table \ref{tab:relatedwork} to make  a comparison with our work. As can be seen, in comparison to these studies, (i) we have automated the implementation of hybrid cloud at the network level using Terraform as an automation tool and WireGuard as a secure, robust, and zero-cost VPN, and (ii) evaluated impact of cloud bursting on the performance of distributed databases. 

\begin{figure*}[htp]
  \centering
  \subfloat[On-demand]{\label{figur:ondemand}\includegraphics[width=0.5\textwidth]{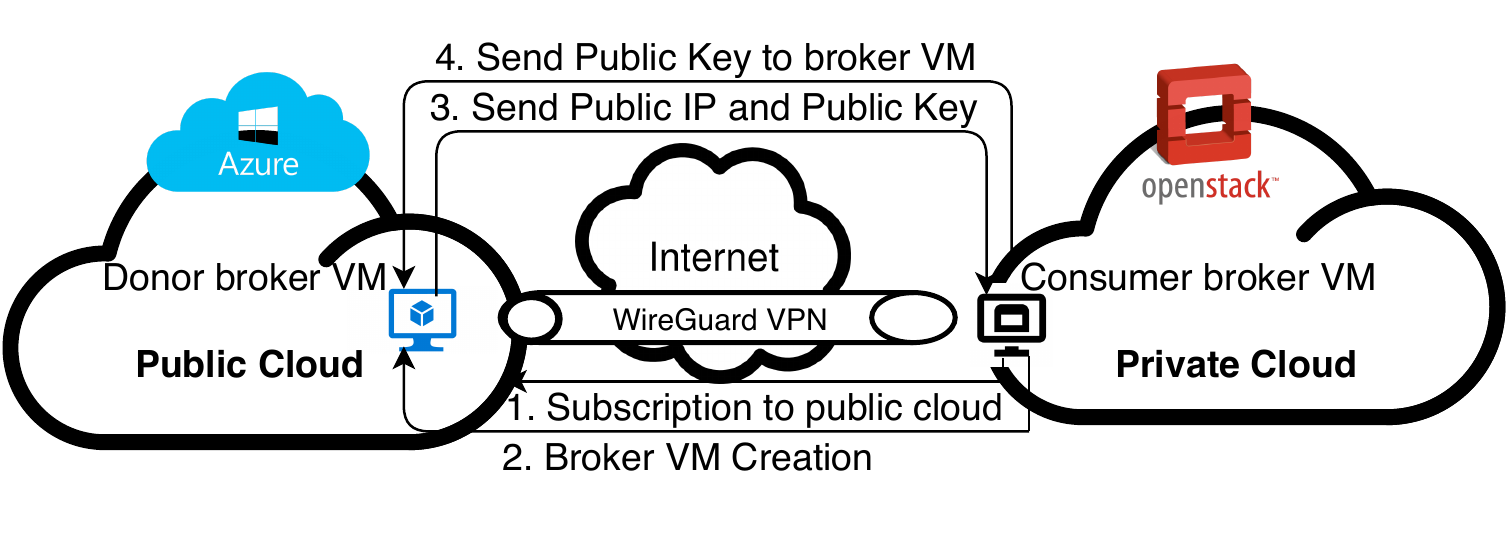}}\hfill
  \subfloat[Fragmented]{\label{figur:federa}\includegraphics[width=0.5\textwidth]{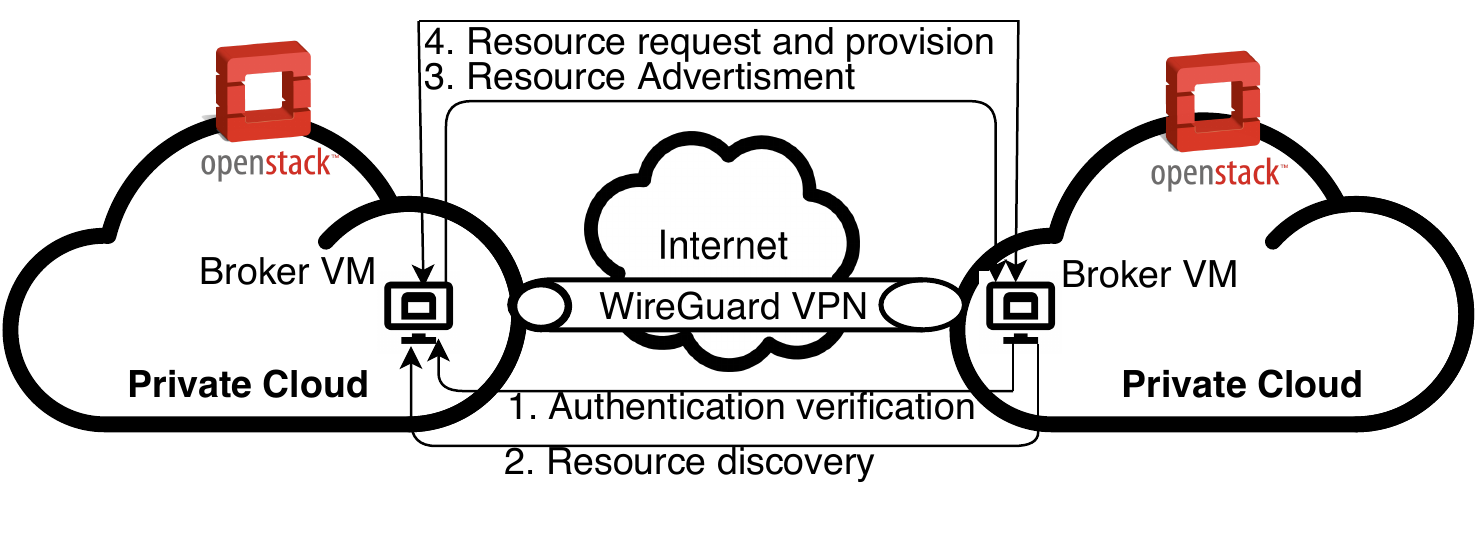}}\par
  \subfloat[Collaborative]{\label{figur:colla}\includegraphics[width=1\textwidth]{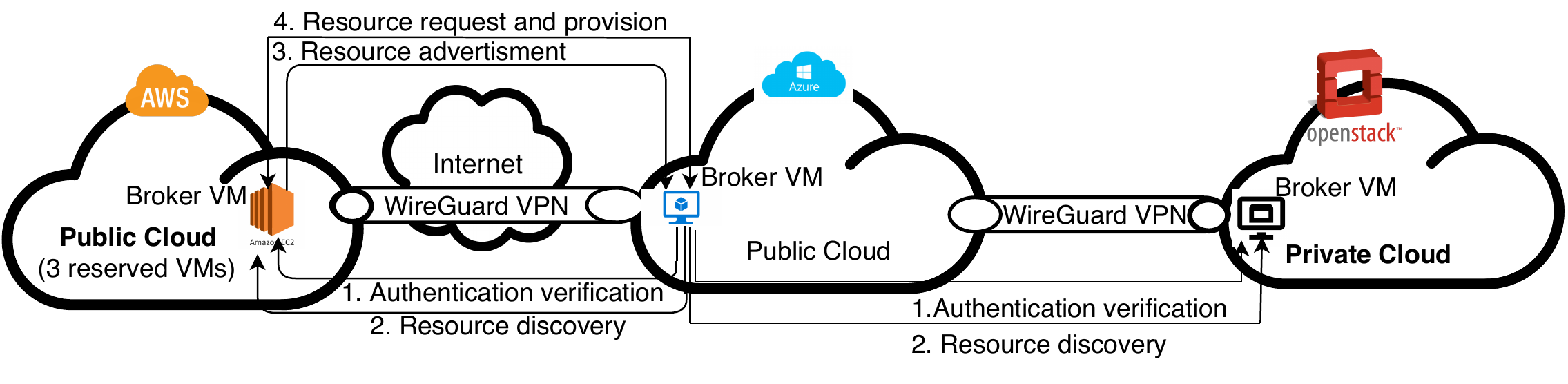}}
  \caption{Different usage models of hybrid clouds. (a) On-demand, (b) Fragmanted, and (c) Collaborative.}
\label{fig:usagemodel}
\end{figure*}

\section {Hybrid Cloud Usage Models}
%In this section, we first propose three usage models of hybrid cloud and then discusses how the common usage model is implemented.
The usage models of the hybrid cloud define resource sharing pattern of  private and public clouds. The connection is influenced by the authentication process and the amount of shared resources in a hybrid cloud. In our models,  a \textit{consumer} cloud requests resources and a \textit{donor} cloud shares its resources with the consumer cloud.
Obviously, consumer and donor clouds can be either a private cloud or a public cloud. In the following, we propose different usage models of  hybrid clouds.

\subsection{On-Demand Usage Model}
This model is known as \textit{cloud bursting } and makes a secure connection between a private cloud and a public cloud, where  the private  and public clouds we respectively call \textit{consumer} and \textit{donor}. When a data-intensive application running on a private cloud and requiring a resource (e.g., a VM instance) the application provider subscribes to a public cloud and creates it in the public cloud using either automation tools such as Terraform\footnote{Terraform: \url{https://www.terraform.io/}}, Ansible\footnote{Ansible: \url{https://www.ansible.com/}}, or APIs provided by the public cloud. After that, the private cloud makes a connection with the public cloud through WireGuard.  Fig. \ref{figur:ondemand}  shows a summary of collaboration between two clouds in three steps: subscription to a public cloud, creation of resource, and connection between private and public clouds by exchanging public keys and VM's public IP in the public cloud. This model can influence the shared resources in terms of security aspect since the data out-sourced and a secure connection  is essential to transfer insensitive data. In contrast, it does not impede the amount of resources to share because public clouds offer an illusion of infinite resources.    

\subsection{Fragmented Usage Model}
This model makes a connection between two or more private clouds and is suitable for mission-critical systems that possess a private cloud. The amount of resources depends on  a sub-system size, where in most cases the smaller sub-system demands resources from the bigger one. As shown in Fig. \ref{figur:federa},  the process of deploying resources across two private clouds consists of authentication verification
between two clouds, dynamic resource discovery, advertisement of resources by donor cloud, resource request and provision.

 In this model of usage, private clouds deploying by different sub-systems can be pre-configured in network connection; as a result it faces less security concerns. The  exploitation of such usage model in a mobile environment leads to as sub-systems/devices are approaching more, the amount of resources available is more. In contrast, as they are moving away the amount of resources available is less. Hence, a challenging problem of this environment is to share resources on-the-fly between different sub-systems/devices due to their mobility. This implies that, in contrast to the previous model, we need to consider connection and disconnection of network and thus dynamic discovery of resources is a must. 
 
 \subsection{Collaborative Usage Model}
This model makes a connection between different models of clouds (i.e., private and public) authorized under a single or multiple collaborating organization. As depicted in Fig. \ref{figur:colla}, an organization provides resources from different cloud providers (e.g., Microsoft Azure, AWS, and Google) as well as private clouds. We may require making network connections between two public clouds, two private clouds, or one private and one public cloud. In collaborative usage model, the resource sharing between two public clouds occurs when the reserved resources are free in the  public cloud while the organization requires resources more than that reserved in the consumer cloud. It is worth noting that public providers offer reserved computation and database resources in terms of one and three years. These resources are cheaper than the on-demand resources but can be wasted if not used in the given time . This encourages different parts of an organization to share their reserved resources between co-workers. In this model, the connection between private and public clouds can be a specific case of the first model while the amount of resource providing by the public cloud is limited.  In summary, the collaborative usage model inherits limitations of the first two models, i.e., security concerns and constrained resources, and in contrast to the second usage model it is a static model.  Table \ref{tab:usagemodel} summarizes the discussed usage models of hybrid clouds in purpose, key challenges, optimization domains, features, and applicability.

\begin{table*}[t]
	% \begin{threeparttable}
	\caption{A comparison of different usage models of hybrid Clouds}\label{tab:usagemodel}
	\centering
	%\vspace{-3mm}
	\tiny
	\begin{tabular}{p{2cm} p{3.1cm}p{4.1cm}p{4cm}}
		%\centering
		\hline
		              &On-demand  &Fragmented &Collaborative \\\hline\hline
		Purpose                 
		     &To gain temporary access to external resources  
		     &Dynamically discover and use shared resources based on geographic location and network reachability 
		     &To use shared pool of resources within a single or multiple collaborating organization \\\hline
		Key challenges      
		        &   Data privacy &Abrupt resource disconnection & Multitude of incompatible clouds \\
                &   Secure communication & Automated resource discovery &Secure Communication\\
                &  	Network routing &  	Network routing &  	Network routing\\
		        & & Secure communication &     \\\hline
		Optimization domains    & Performance &Latency &Performance      \\
		                        & Latency &Bandwidth &Latency     \\
		                        & Bandwidth &- &Bandwidth     \\
		                        & Monetary cost &- &-     \\\hline
		Features                & No upfront cost &Up-front cost &Depend on combination   type of clouds     \\
		                        & No data-lock in & Data lock-in &-\\
		                        & A variety of resource types & Limited resource type &-\\\hline
		Applicability           & Business organizations  &Critical-mission systems &Depend on combination of cloud types    \\\hline

	\end{tabular}
	%\end{threeparttable}
	\vspace{-5mm}
\end{table*}

\section{An Automated Implementation of Hybrid Cloud}
Among the usage models of hybrid clouds discussed above, we select on-demand usage model as one of  the most conventional deployment of hybrid clouds\footnote{To implement two other usage models, it is required to add two key components on top of the on-demand usage  model. The first component is to implement a database to include used and free resources across all cloud datacenters participating in the hybrid cloud. This database allows to make queries to ensure the availability of required resources. The second component is to add a mechanism to verify authentication between cloud datacenters.}. Two challenging issues relating to the  implementation of this model is discussed below.
%within on-demand usage model (in general hybrid cloud) is how to make a secure connection between two clouds, and how to deploy resources across two different clouds. In this section, we discuss how these challenges are addressed and what criteria are considered in  selection of technologies for an automated implementation of hybrid cloud. 

\textbf{Connectivity issues in hybrid clouds:}
A secure connectivity\footnote{A secure connection makes encrypted data by security protocols to guarantee the security of data flow between VMs locating across private and public clouds.} is one of the integral parts of the deployment of a hybrid cloud since  virtual machines (VMs) are scattered across networks with different administration domains. One possible solution  is the allocation of public IP addresses to the deployed VMs in the private cloud. However, allocation of public IPs to VMs in a  private cloud is  waste of resources as if there is no limitation in allocation of public IPs   by an IT business. Furthermore, VMs in a private cloud usually operate behind organizational firewalls and network address translation (NAT), which hinder developers to make connections via public IPs.

However, network virtualization techniques provide an alternative  connection via public IPs. A Virtual Private Network (VPN)  is a secure way to build an overlay network over  Internet and to make a secure connection between private and pubic clouds. For this purpose,  the well-known cloud providers offer VPNs like Amazon VPC\footnote{Amazon VPN:\url{https://aws.amazon.com/vpc/}}, Google VPN\footnote{Google VPN:\url{https://cloud.google.com/vpc/}}, and Microsoft Azure VNet\footnote{Azure VPN:\url{https://azure.microsoft.com/en-au/services/vpn-gateway/}}. Since we implement a hybrid cloud  over Azure and OpenStack clouds, Azure VPN can be used for the hybrid cloud connection. 

Microsoft Azure offers three VPN options.\footnote{Point-to-Point (P2P) creates a secure connection from an individual  VM in the private cloud to the Azure VNet, while VNet-to-VNet (V2V) makes a secure connection between two VNets in the same or different regions in Azure cloud.} Site-to-Site(S2S) meets the requirements of a hybrid cloud construction and  needs  a VPN gateway  with an assigned Public IP over the public cloud side. However, we  observed that Azure VPN gateway connection stops after about two and a half hours if remains inactive. Thus, it requires a mechanism to wake up regularly with the help of third-party scripts or built-in Azure mechanisms. In addition, Azure charges users based on the amount of time that the VPN gateway is provisioned. This charge is applied even if no data is sent and received through VPN. 

\begin{figure*}[h!]
  \centering
  
  \subfloat[]{\label{figur:azopvpn}\includegraphics[width=0.7\textwidth]{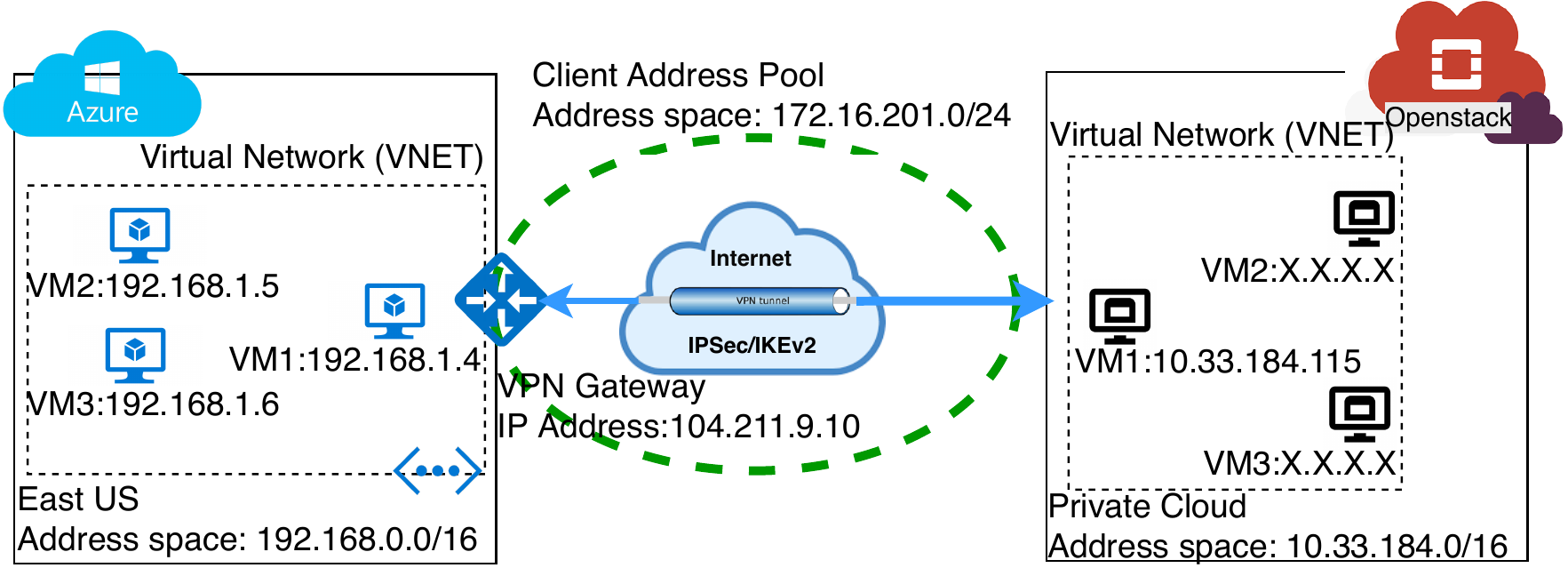}}\\
 \subfloat[]{\label{figur:azopwi}\includegraphics[width=0.7\textwidth]{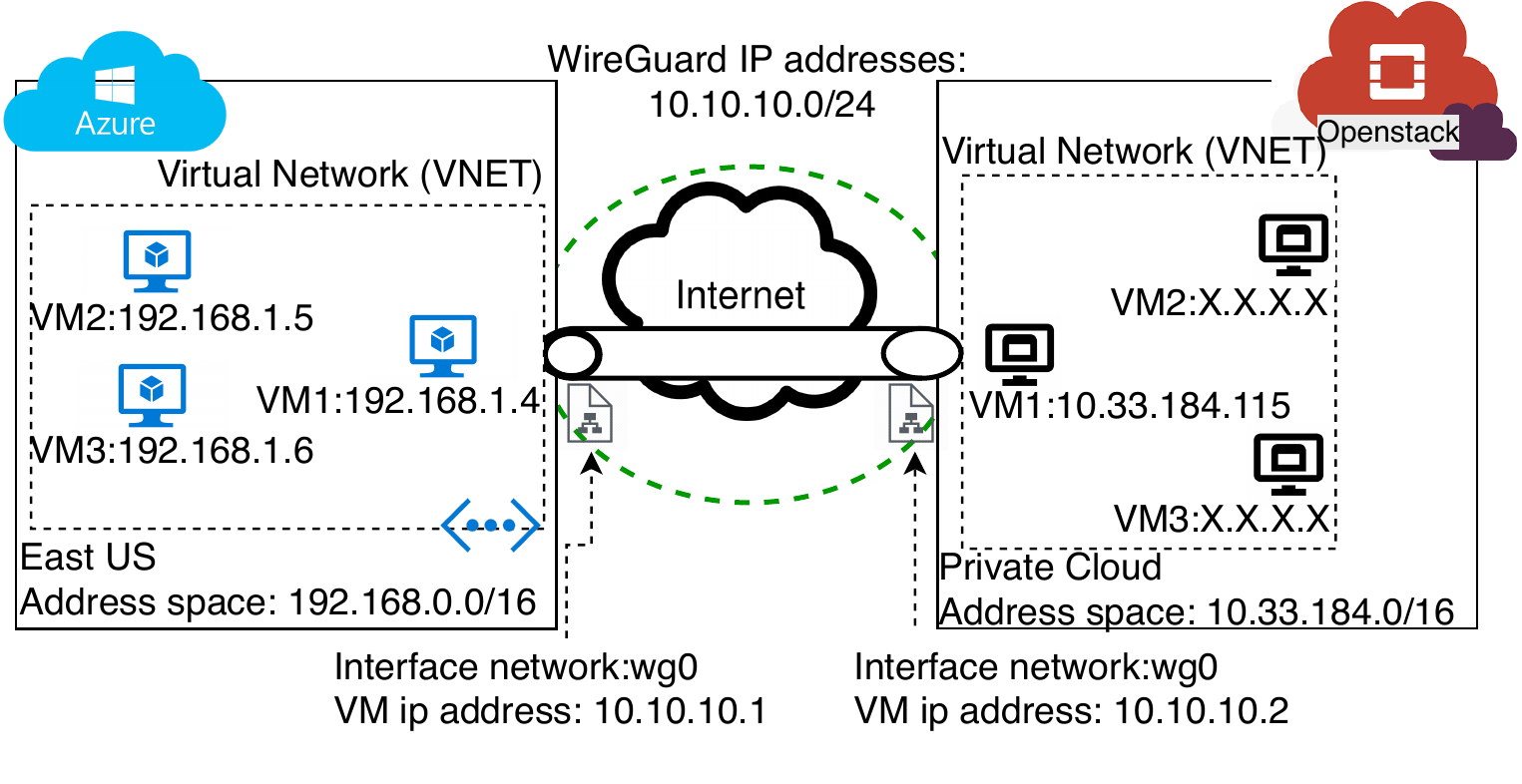}}
  %\subfloat[]{\label{figur:colla}\includegraphics[width=0.5\textwidth]{Fig/collusmo.pdf}}
  \caption{Connection between OpenStack and Azure using (a) WireGuard and (b) Azure VPN Gateway}
\label{fig:hybconnec}
\end{figure*}

To avoid these issues, we deploy  WireGuard as  Linux kernel-based  VPN recently released. It is designed to be easy to use while providing a robust and secure connection between parties over a network interface encrypted with public key authentication. 
%Unlike most VPNs, no topology is enforced so different configurations can be achieved by manipulating the surrounding networking configuration. 
WireGuard, in contrast to Azure VPN,  gives a virtual interface, for example wg0, which can be managed using the standard ip(8) and ifconfig(8) utilities. After configuring the interface wg0 with a private key and public keys of peers with whom it will connect securely, the tunnel between peers becomes operational. This makes WireGuard easier to use and simple compared to IPSec used by Azure VPN gateway. Furthermore, there is no need of VPN gateway and a public-facing IP address in on-premise infrastructure. Fig. \ref{fig:hybconnec} illustrates the deployment of Azure VPN and Wiregurad to build a hybrid cloud. 

The exploitation of WireGuard instead of Azure VPN brings the following benefits. \textit{Security}: Wireguard authenticates peers through exchanging public keys and encrypts  data independent of any vendor/organization. While Azure VPN encrypts data which is under control of a vendor. \textit{Reliability in connection}: WireGuard makes a persistent connection between clouds even without sending and receiving data between two VMs as conducted experiments for 72 hours; while Azure VPN does not. \textit{Cost:} WireGuard offers zero-cost services while Azure VPN charges users. \textit{Ping Time and Throughput}: WireGuard outperforms IPSec protocol \cite{Dhall2012}  (exploited by Azure VPN) in both ping time (i.e., response time) and throughput \cite{Donenfeld2018}. \textit{Interportability}: WireGuard easily provides interportability  between any cloud provider through creating Wireguard configuration file. 
%in two VM connected that should be connected to each other. 
Thus, we use WireGuard to implement the hybrid cloud. 

\textbf{Deployment of Infrastructure Resources:} There are several ways to define, preview, and deploy cloud infrastructure resources. Public cloud providers support several embedded tools like  command-line interface (CLI), Power Shell, and Portal web services that enable the management of  limited resources with static properties. To break these limitations,  Infrastructure as Code (IAC) like Terraform\footnote{Terraform:\url{https://www.terraform.io}} support resource deployment and configuration in a dynamic way. Terraform is strictly declarative language and provides a readable and clear syntax for software developers and makes it more desirable to be used in the configuration of infrastructure resources. Hence, we select it to implement resource pooling across OpenStack and Microsoft Azure.

\begin{figure}[t]
	\centering
	\includegraphics[width=1\columnwidth]{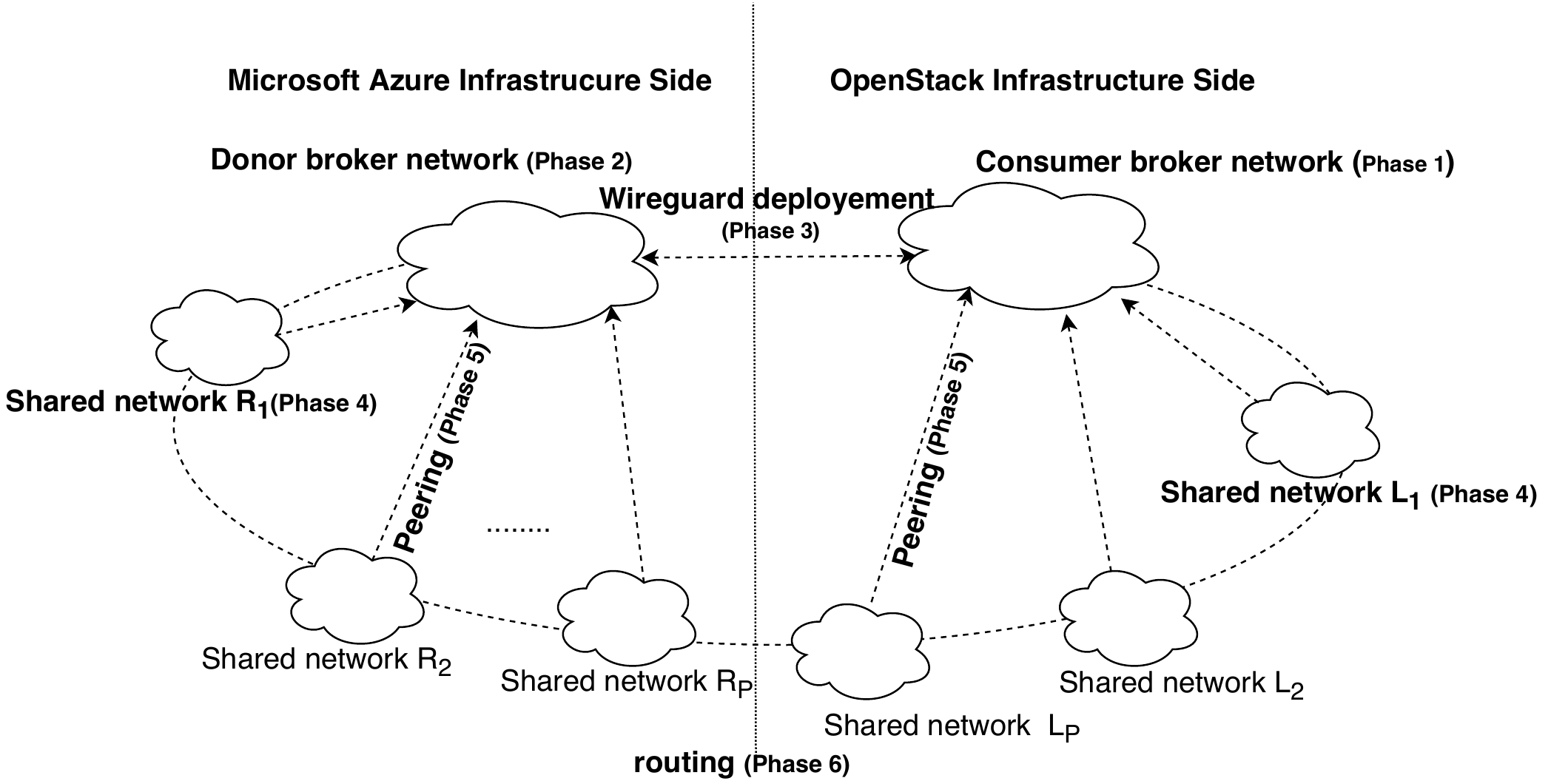}
	\vspace{-1mm}
	\caption{Different Phases of HybOPAZ Implementation}
	\label{fig:phases}
	\vspace{-4mm}
\end{figure}

\begin{figure}[t]
	\centering
	\includegraphics[width=0.5\columnwidth]{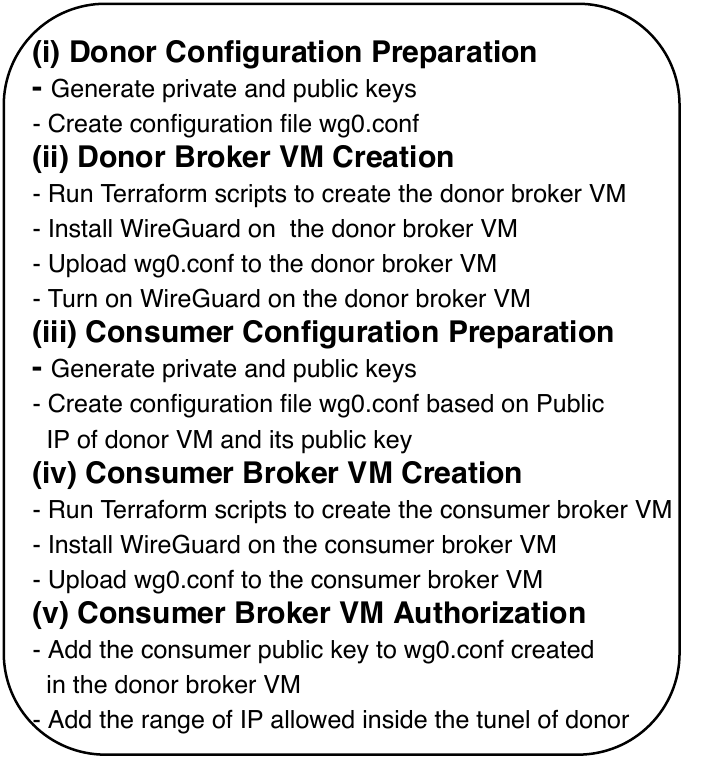}
	\vspace{-1mm}
	\caption{A Procedural Step-by-Step of Donor and Consumer Broker VMs Configuration Preparation, Creation, and Authorization}
	\label{fig:modseq}
	\vspace{-4mm}
\end{figure}

\textbf{Phases of Hybrid Cloud Implementation:} We discuss step-by-step the implementation of hybrid cloud over OpenStack and Microsoft Azure (HybOPAZ) based on on-demand usage model using WireGuard and Terraform. Fig. \ref{fig:phases} shows  the phases of HybOPAZ implementation in which initially a consumer broker network is created within OpenStack (Phase 1). To expand workload to the public cloud in the case of workload spike, we need to create a donor broker network on Azure and connect it to the consumer broker through WireGuard (Phases 2-3). Based on the requirements, we enable HybOPAZ to expand shared networks/subnetworks (phase 4) that can be peered (phase 5) to the local networks. Finally, Phase 6 enables data routing across broker, local and shared networks for HybOPAZ. We discuss the implementation of phases below.

%We implement  phases in separate modules to ease the implementation of infrastructure deployment, networking, and data routing. The modularity allows the application providers/organizations to run separate modules based on their requirement. We now discuss how to automate  different steps as summarized in Fig. \ref{fig:modseq}.

\begin{figure}[t]
	\centering
	\includegraphics[width=1\columnwidth]{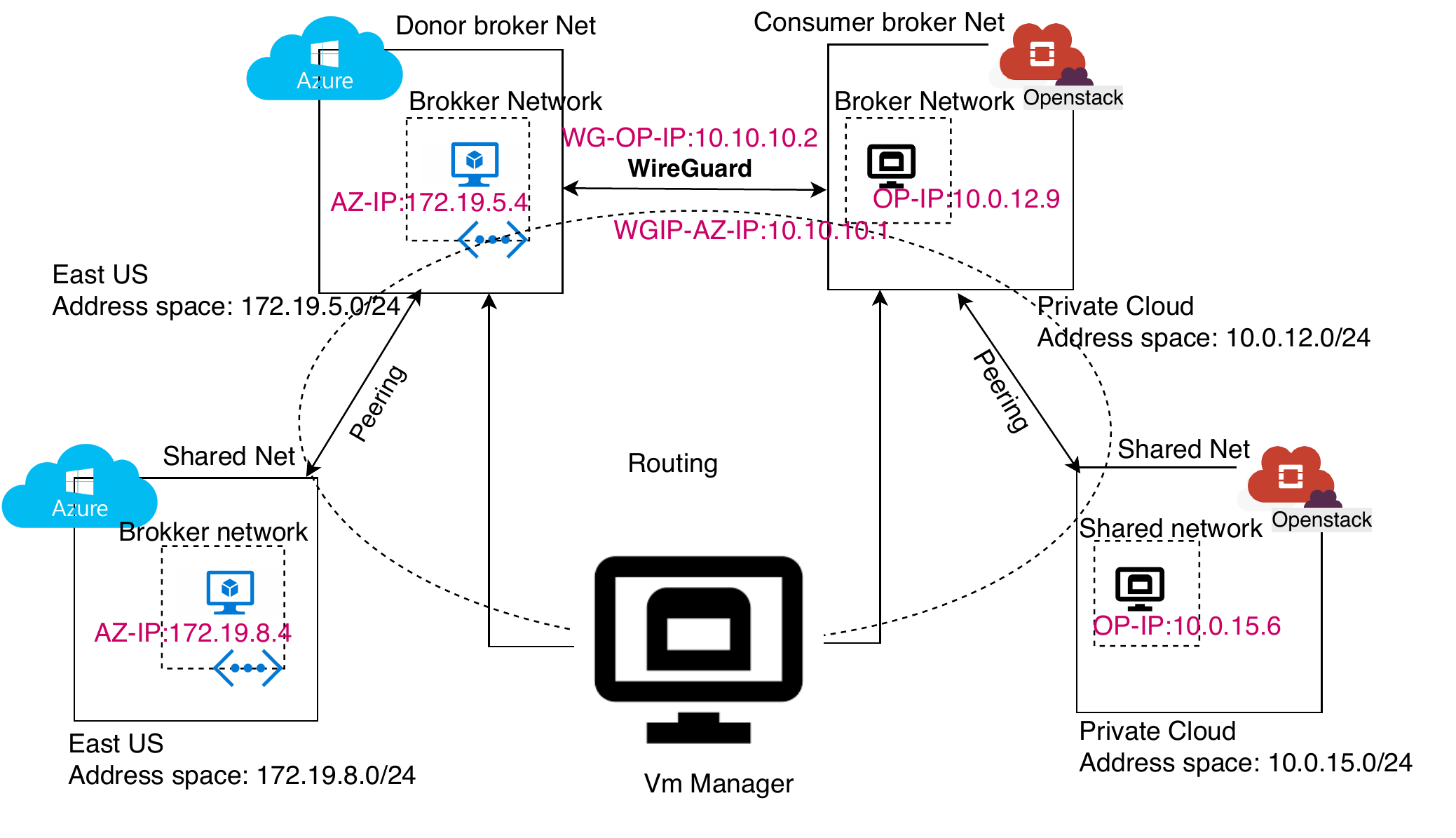}
	\vspace{-1mm}
	\caption{Implementation of HybOPAZ using WireGuard}
	\label{fig:impHyOPAZ}
	\vspace{-4mm}
\end{figure}

%\begin{figure}[t]
%	 \centering
%	\includegraphics[width=0.5\columnwidth]{Fig/attpro.pdf}
%	\vspace{-1mm}
%	\caption{Script of authorization and deauthorization of consumer broker VM}
%	\label{fig:authconvm}
%	\vspace{-2mm}
%\end{figure}

\textbf{Step 1}: This step consists of phases 1-3 of HybOPAZ implementation including creation of consumer and donor broker VMs, and WireGurd configuration between them. As depicted in Fig. \ref{fig:impHyOPAZ},  we deployed a VM manger in OpenStack to control and store private and public keys required for a secure connection between broker VMs.  We wrote several Terraform  and shell scripts for the following modules (see Fig. \ref{fig:modseq}) to automate resource sharing between two clouds.

 (i) \textit{Donor configuration preparation}: This module creates public and private keys via key generation mechanism provided by WireGuard and saves them in the VM manager. Based on these keys, we create WireGuard configuration file wg0  for the donor broker VM in Azure.
 
(ii) \textit{Donor broker VM Creation}: VM manager runs the automation scripts to create a virtual network including the donor broker VM. Then, the terraform script installs Wireguard on it, uploads wg0 to it, and  starts Wireguard services on it.

(iii) \textit{Consumer Configuration Preparation}: This module performs the same task as for the donor broker VM in module (i). Based on the generated keys and  donor broker VM's public key and public IP, we create configuration file for the consumer broker VM in OpenStack.

(iv)\textit{Consumer broker VM creation}:  Like module(ii), VM manger creates the consumer broker VM in OpenStack, then installs Wireguard on it, and finally uploads WireGuard configuration file, created in module(iii), to it. In contrast to the donor broker VM, the consumer broker VM cannot start Wireguard in this step because it is not yet authorized to access the donor broker VM. 

(v) \textit{Consumer broker VM authorization}:  This module allows the consumer broker VM to connect to the donor broker VM through  adding the pubic key of the consumer broker (created in module(i)) and the range of IP addresses allowed inside the tunnel (consumer\_VPN\_IP/32). This authorization functionality is added to Wireguard configuration file in the donor broker VM. %by running authorization script in Fig. \ref{fig:authconvm}. In addition, we write a script that allows to deauthorize a consumer VM from its corresponding donor broker VM via running the script (Fig. \ref{fig:authconvm}) in the VM manager.  

\textbf{Step 2}: This step includes phases 4-6 in the case of workload spikes. In phase 4, the implementation of shared and broker networks is identical in OpenStack. In contrast, this phase is different in Azure since the connection between two subnets with different address spaces via the virtual network peering is provided. Phase 5 connects the shared networks to the broker networks via adding the network interface of shared network (Fig. \ref{fig:impHyOPAZ}) in OpenStack and  virtual network peering in Azure.
Phase 6 implements data routing in OpenStack by adding static rules to both routers in the broker and shared networks. Differently, in Azure, the shared network is initially associated with the routing table of the broker network, and accordingly the routing table is modified with static rules.

\section{Distributed Databases and Workload Setup}
This section initially discusses the selected databases to deploy in the hybrid cloud. Then, it describes the workload for evaluation of these databases.  
\subsection{Distributed Databases}
The main purpose of the automated implementation of hybrid cloud is to evaluate the performance impact of cloud bursting on distributed database systems. This provides an overview of the performance impact of network bottleneck between two datacenter clouds on different distributed database systems. The criteria selection for the NoSQL databases is maturity and high usage in industry. We also selected MySQL as the leading  relational database due to its well-developed and globally exploitation in industry. 
%Our selection of these databases are based on the popularity and maturity. Although there is rich literature on distributed datastores\cite{Mansouri2017a}, in the following we briefly discuss them  to make easy understanding of our comparison in the next section. 
 
 \textbf{MongoDB}: MongoDB is a document-oriented database without single point of failure since in case master replica goes down then the secondary replica is selected as a master. MongoDB supports asynchronous  master-slave replication, meaning that writes are only served by the master replica and reads by any replica. 
 %MongoDB facilitates  application-level encryption for ``data at rest", and uses TLS/SSL to encrypt ``data at motion".

\textbf{ Cassandra:} Apache Cassandra provides a highly available data services with no single point of failure implying all nodes are equal- no concept of master and slave nodes. In Cassandra, each server can handle read and write with different levels of consistency, where ``ONE" is a default level. 
%Cassandra supports optional data encryption for ``data at rest", while it does not offer data encryption for ``data at motion". 
%In our evaluation, we use the recent ??? version with default configuration in data replication, consistency and data distribution. 

%IN our implementation, we use the recent ??? version with default configuration in replication and \textit{eventual} consistency level.   

\textbf{Riak}: Riak is a document-oriented database system and supports master-less replication architecture without single-point of failure. Riak offers tunable consistency level. By default, Riak supports peer-to-peer replication and eventual consistency. 
%Riak does not encrypt ``data at rest", and ``data at motion" through explicit configuration.

\textbf{CouchDB:} Couchdb is a document-oriented database and has  asynchronous master-slave replication and provides local quorum -based consistency. Each database in couchdb is spilt to 8 shards, while other databases leverage a dynamic sharding policy.
%Couchdb does not support ``data at rest" encryption while it is possible using https connection to secure ``data at motion".

\textbf{Redis}: Redis is an in-memory database and supports asynchronous master-slave replication architecture in which each master can handle several slaves and a slave can act as master to other slaves. In fact, Redis supports non-blocking replication on both master- and slave-side, implying that master can serve queries while slaves are synchronizing with master and slave can handle requests through old version of data. 
%Redis does not offer data encryption neither for ``data at rest" and nor for ``data at motion". 
%In our evaluation, we use Redis cluster with version ??? and deploy it with default configuration in replication and consistency level.  

%We use the recent version ??? in our implementation with default configuration for replication and consistency. 

%In our evaluation, we use its cluster version of ??? with eventual consistency by default. 

\textbf{MySQL Cluster}: MySQL Cluster is a distributed and relational database with full SQL support and ACID properties.   It provides shard-nothing clustering and auto-sharding for MySQL database. MySQL Cluster exploits synchronous replication via a two-phase commit to guarantee ACID features.  
MySQL Cluster implemented through Network Database (NDB) is not secure in communication between data nodes in the cluster. Thus, MySQL cluster improves network speed.

We run all these databases based on the default settings as summarized in  Table \ref{tab:distributedDBs}. As can be seen, some of those databases do not support data encryption for ``data at rest'' and ``data at motion" and can take benefit from Wiregurad to encrypt data.

%but does not support ``data at motion" encryption. In contrast, it offers data encryption in disk and database level. 
%In our evaluation, we use the recent version of MySQL Cluster (???) with default replication and consistency. 

\subsection{Workload setup}
To evaluate cloud bursting at the database level, we leverage YCSB \cite{Cooper2010} workload that is suitable for store and retrieve data from distributed databases. YCSB consists of \textit{ client component} that generates workload and \textit{core workload} that is a set of queries to be executed by the client component.  This workload allows to define a configurable distribution of CRUD (create, read, update and delete) operations on data set. Each record in the data set consists of 10 fields, each with a length of 8 bytes. Thus, a record in the data set is 80 byte. 

As shown in Table \ref{tab:coreworkload}, YCSB workload consists of one or two atomic operations and includes three phases: \textit{flush}, \textit{load}, and \textit{run}. Flush phase allows to make clear database from the data already stored. Load phase generates 10000 records of  random data and writes in the database. Run phase executes 1000 atomic operations on the written data in database. All these phases run on the local OpenStack to keep validation of on-demand usage model of hybrid cloud.

\begin{table}[t]
	% \begin{threeparttable}
	\caption{Core workload in YCSB}\label{tab:coreworkload}
	\centering
	%\vspace{-3mm}
	\begin{tabular}{p{2.7cm} p{4.5cm} p{5.5cm}}
		%\centering
		\hline
		Workload type &         Operations &Label    \\\hline\hline
		Workload A     & 50\% Read + 50\% Update & Read-intensive  \\
		Workload B     & 95\% Read + 5\% Update  & Write-intensive  \\
		Workload C     & 100\% Read              & Read-only    \\\hline
		Workload D     &95\% Read + 5\% Insert   & Read-latest\\
		Workload E     &95\% Scan + 5\% Insert   & Scan\\
		Workload F     &50\% Read + 50\% update        &Read-Modify-Write(RMW) \\\hline
	\end{tabular}
	%\end{threeparttable}
	\vspace{-5mm}
\end{table}

\section{Performance Evaluation }
 We delineate the details of hybrid cloud infrastructure resources and discusse experimental evaluation and practical findings of cloud bursting at the database level.
 
\subsection{Hybrid Cloud configuration}
We used two clusters for our evaluation based on the HybOPAZ architecture. One group of nodes is deployed in local OpenStack (version:2.3.1) and consists of VM instances with 1 core CPUs, 2 GiB of RAM, and 10 GB disk. Other group of nodes is created in Azure cloud datacenter with Standard\_B1ms instances (1 vCPU, 2 GiB RAM, and 4 GiB SSD storage. In all experiments\footnote{The key purpose of these experiments with such hardware resources is to compare the performance of these databases in the same infrastructure resources. However, we plan to investigate vertical scalability (increasing capacity of hardware) and horizontal scalabilty (adding more VMs to the pool of resources) for distributed databases across hybrid clouds.}, we deployed 8 total nodes in both groups so that \textit{n} nodes are exploited in the local OpenStack, and \textit{8-n} nodes are burst into the Microsoft Azure cloud in East US region. In our experiments, we considered all permutation of 8 nodes that can be burst into the public cloud. Thus, we have  setups of (8,0), (7,1), (6,2), ..., (2,6) and (1,7) across both clusters, where the first and second member of each pair is the number of VMs, respectively, deployed in OpenStack and Azure. 

We installed databases and configured them on both groups as a single cluster based on the default settings as summarized in Table \ref{tab:distributedDBs}. For this purpose, we deployed two phase scripts: installation and cluster configuration. For MySQL, cluster configuration is slightly different in comparison with NoSQL databases since it requires three types of nodes: \textit{Cluster manager node}, \textit{data nodes}, and \textit{MySQL Server node}. We run cluster manager node and MySQL server node on the same VM instance in OpenStack, and data nodes across both clusters. 

With the help of Terraform, we automated deployment and destruction of infrastructure resources across OpenStack and Microsoft Azure. We also leverage Terraform to run configuration files in order to install and build a cluster of distributed databases across the exploited infrastructure. Such automation of implementation allows us to consistently recreate the clusters of distributed database nodes based on the required VMs number, VMs flavor, VMs region, network features, etc.

%As shown in Table \ref{tab:distributedDBs}, for Mongo, Redis, Riak, and CouchDB, we   consider only a configuration file because all nodes are tread the same.

\begin{table}[t]
	 \begin{threeparttable}
	\caption{Default setting for six distributed databases}\label{tab:distributedDBs}
	\centering
	%\vspace{-3mm}
	\begin{tabular}{p{2cm} p{2cm} p{2.7cm} p{1.7cm} p{1.8cm} p{1.5cm}}
		%\centering
		\hline
		Database        &Replica number  &Consistency &Data at rest &Data at motion &Version    \\\hline\hline
		MongoDB           &Full    &Eventual &Yes &Yes   & 3.4  \\
		Cassandra       &3       &Quorum  &Yes &No     & 3.11  \\
		Redis           &Full    &Eventual &No &No  &5.0 \\\hline
		Riak            &3      &Eventual  &No &Yes   &2.2.3 \\
		CouchDB         &3      &Local Quorum &No &Yes  & 2.2.0\\
		MySQL    &2       &Strong   &Yes &No  &7.6.6\\\hline
	\end{tabular}
	\begin{tablenotes}
      \small
      \item Columns ``data ate rest'' and ``data at motion'' indicate whether or not a database system respectively supports data encryption at storage level and network level between storage nodes.
   \end{tablenotes}
   \end{threeparttable}
	\vspace{-3mm}
\end{table}

\subsection {Experimental Results}
In this section, we answer to this research question: \textit{How effective is the Wide Area Network (WAN) cloud bursting at the database level?}. To this end, we initially evaluate HybOPAz  in terms of latency and bandwidth between different sub-networks to better understand the limitations of distributed databases deployment across hybrid clouds. Then, we report the results of such deployment in respect to throughput, latency of read and write operations, and the error percentage for operations performed on distributed databases.

\begin{figure}[htbp]
\begin{minipage}[t]{0.48\linewidth}
    \includegraphics[width=\linewidth]{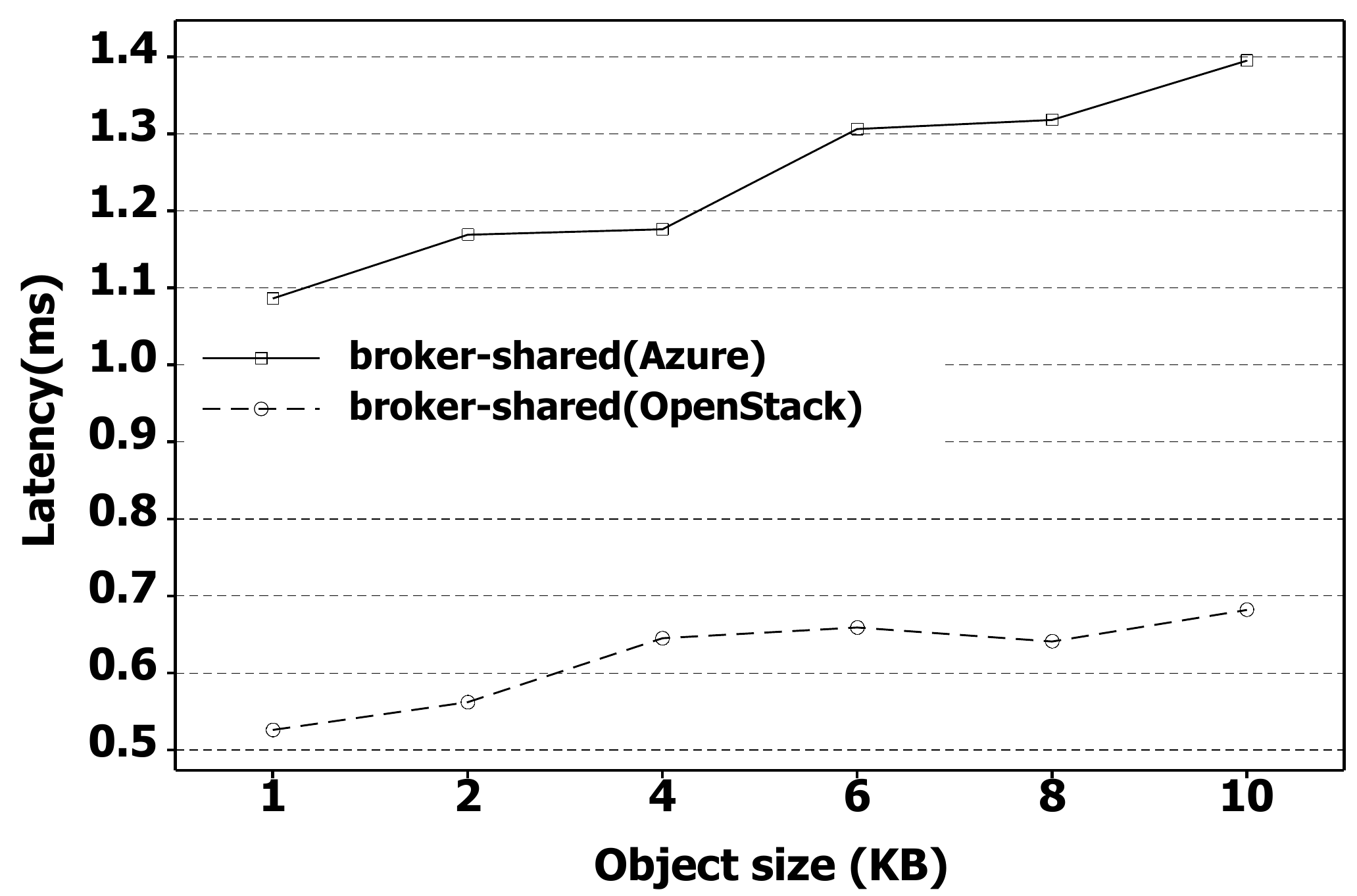}
    \vspace{-15pt}
    \caption{Latency between broker and shared sub-networks in OpenStack and Azure}
    \label{figur:az-op}
\end{minipage}%
    \hfill%
\begin{minipage}[t]{0.48\linewidth}
    \includegraphics[width=\linewidth]{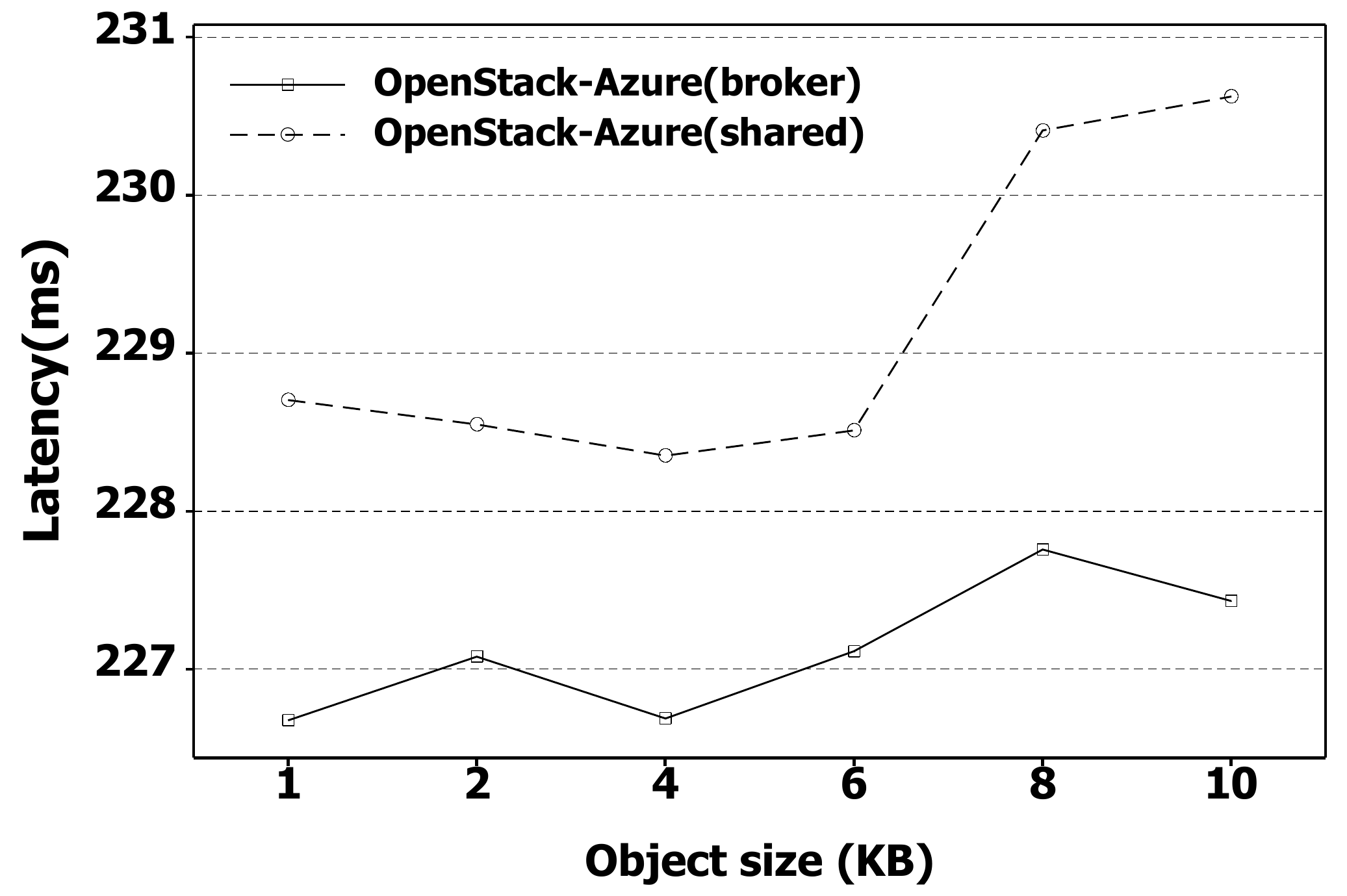}
    \vspace{-15pt}
    \caption{Latency between two brokers and two shared sub-networks across OpenStack and Azure}
    \label{figur:op-az-across}
    %\vspace{-5mm}
\end{minipage} 
\end{figure}

As shown in Fig. \ref{figur:az-op}, latency between broker and shared sub-networks varies between 0.5-0.7ms in OpenStack and 1.1-1.4ms in Azure to transfer 1000 data packets. In contrast, as depicted in Fig. \ref{figur:op-az-across}, these values respectively increase by  220ms-230ms  and 228.5ms-230.5ms for two broker and two shared networks in HybOPAZ. Thus, as expected, the latency between two sub-networks across clouds is significantly high and unpredictable. This is confirmed by our experiment, with the help of MTR utility\footnote{MTR:https://www.bitwizard.nl/mtr/.}. MTR shows that transfer of data packets between the two shared networks requires to pass 24 hops;  12 of which are in Australia and require 12ms to pass data packets, with the remaining hops outside of Australia and requiring 218ms to transfer data packets. 

Furthermore, we measure the bandwidth between two shared networks in HynOPAZ through running IPerf3\footnote{iPerf3: https://iperf.fr/.} for 10 minutes.  Results show that the bandwidth for download and upload data between these two networks respectively reaches  11191 and  1009 Kbyte/sec. This  demonstrates another restriction for transferring data between two clouds. In respect to these limitations, we evaluate cloud bursting at the database level.

\begin{figure*}[h!]
  \centering
  \subfloat[Mongo]{\label{figur:mongo-thr}\includegraphics[width=0.5\textwidth]{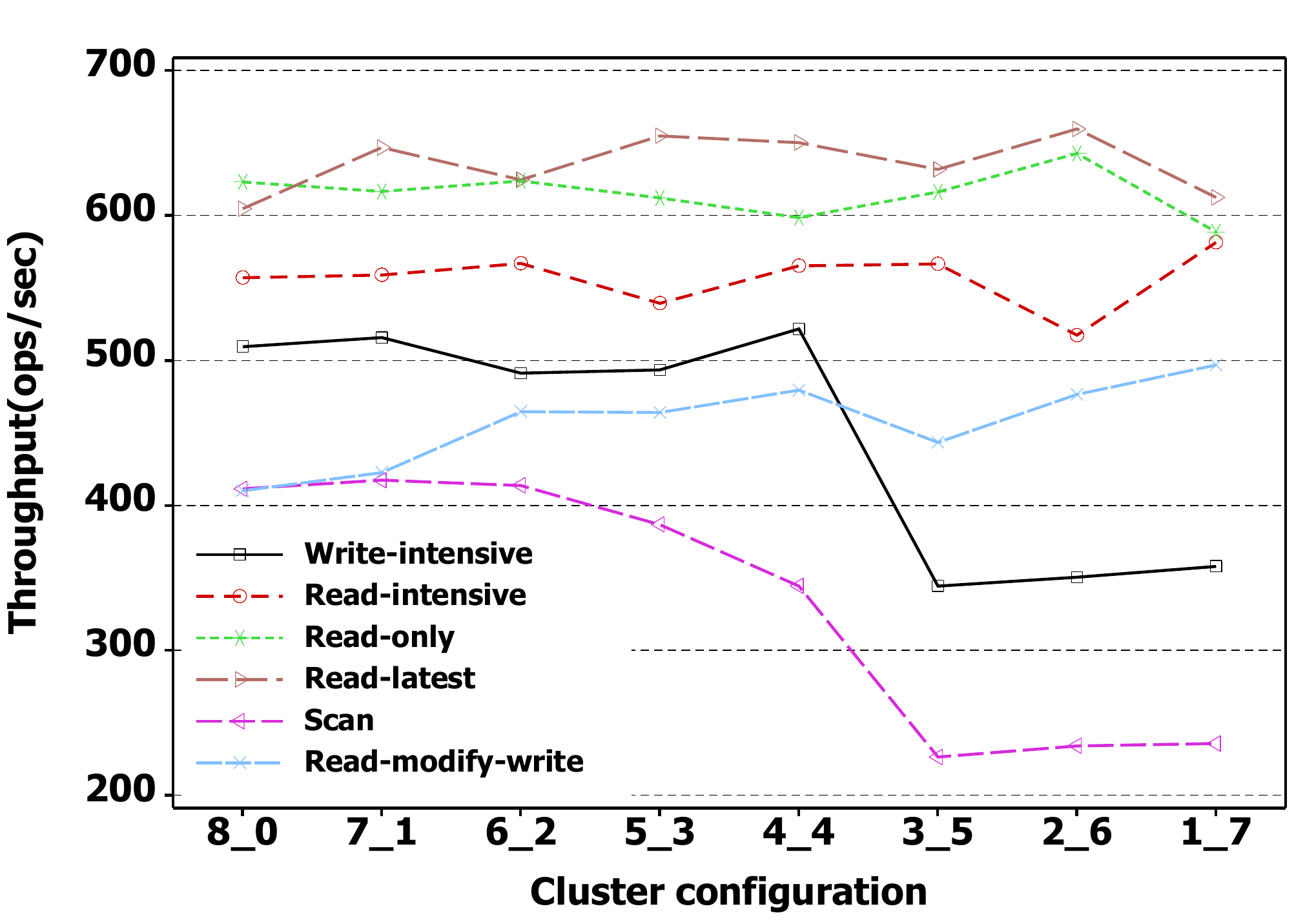}}
  \subfloat[Cassandra]{\label{figur:cassandra-thr}\includegraphics[width=0.5\textwidth]{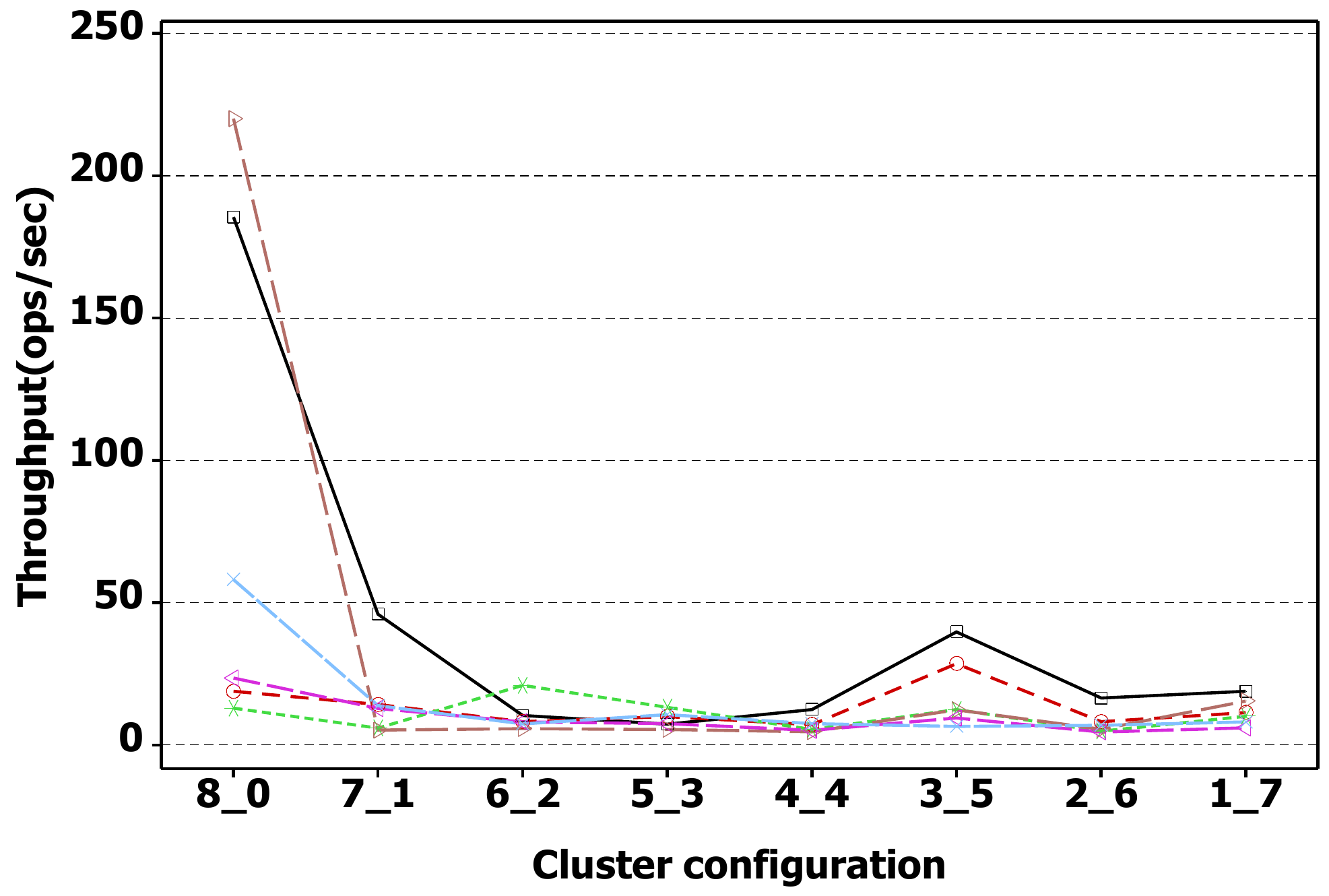}}\\
  \subfloat[Riak]{\label{figur:riak-thr}\includegraphics[width=0.5\textwidth]{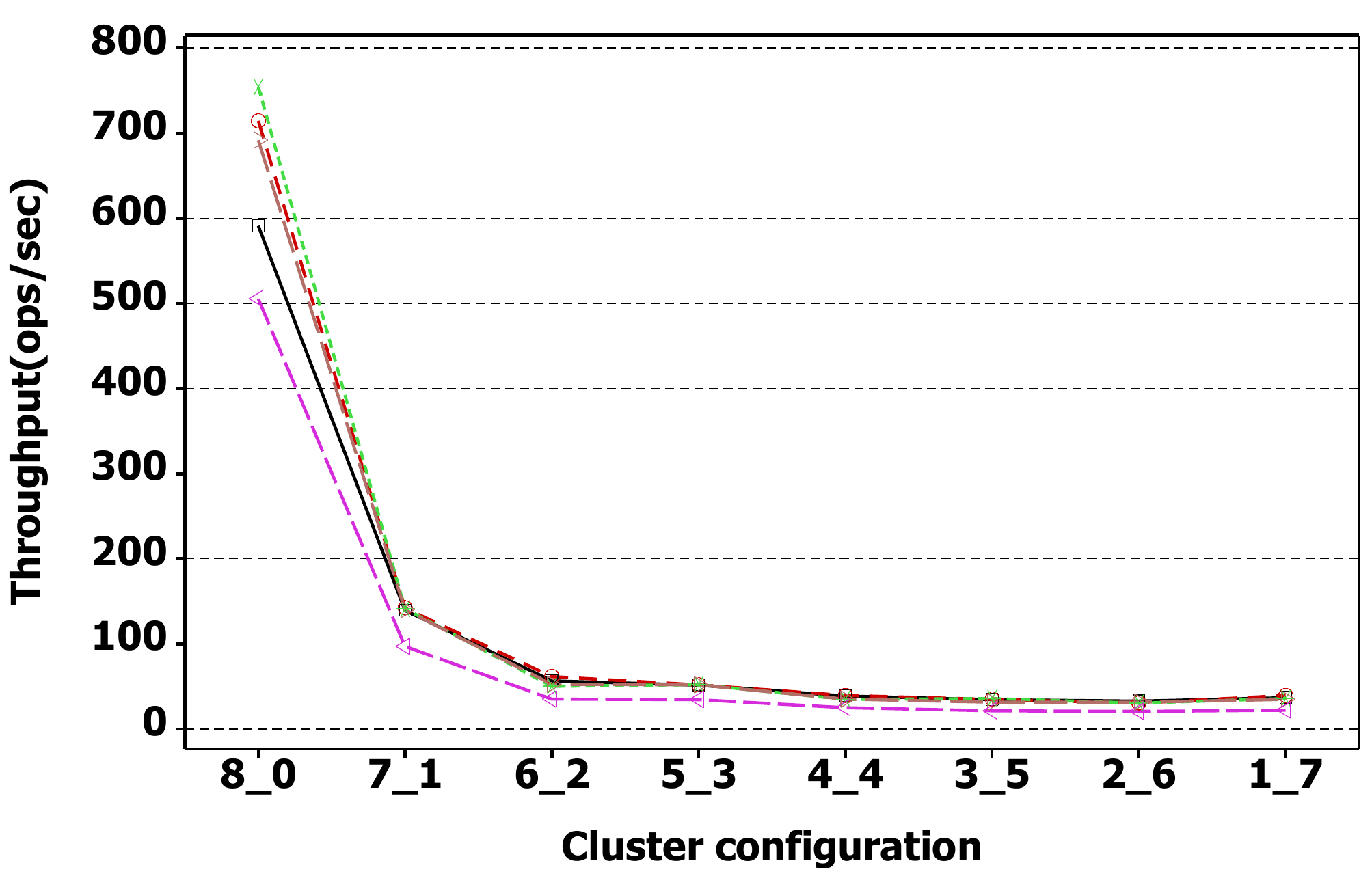}}
  \subfloat[Couchdb]{\label{figur:couchdb-thr}\includegraphics[width=0.5\textwidth]{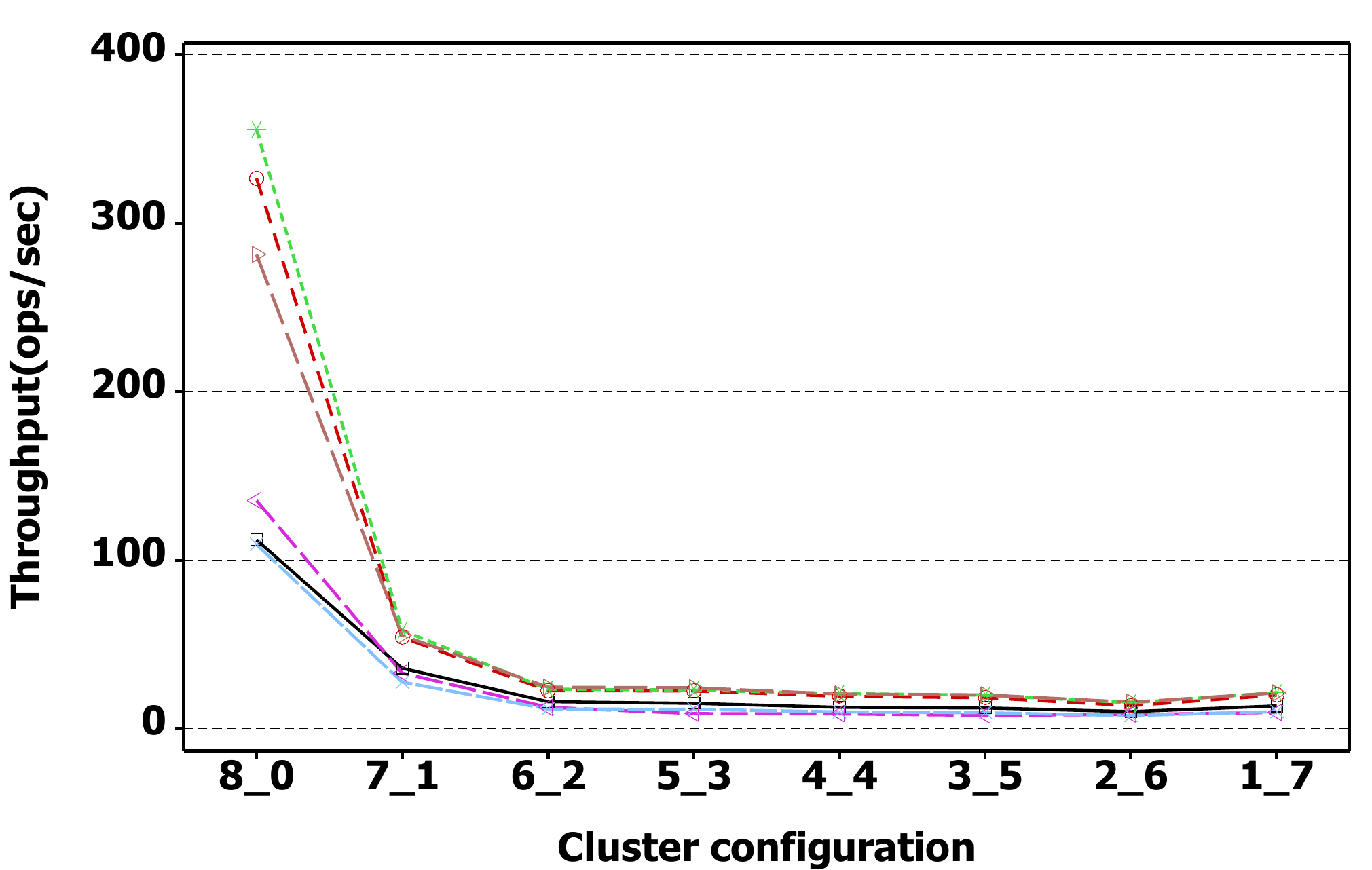}}\\
  \subfloat[Redis]{\label{figur:redis-thr}\includegraphics[width=0.5\textwidth]{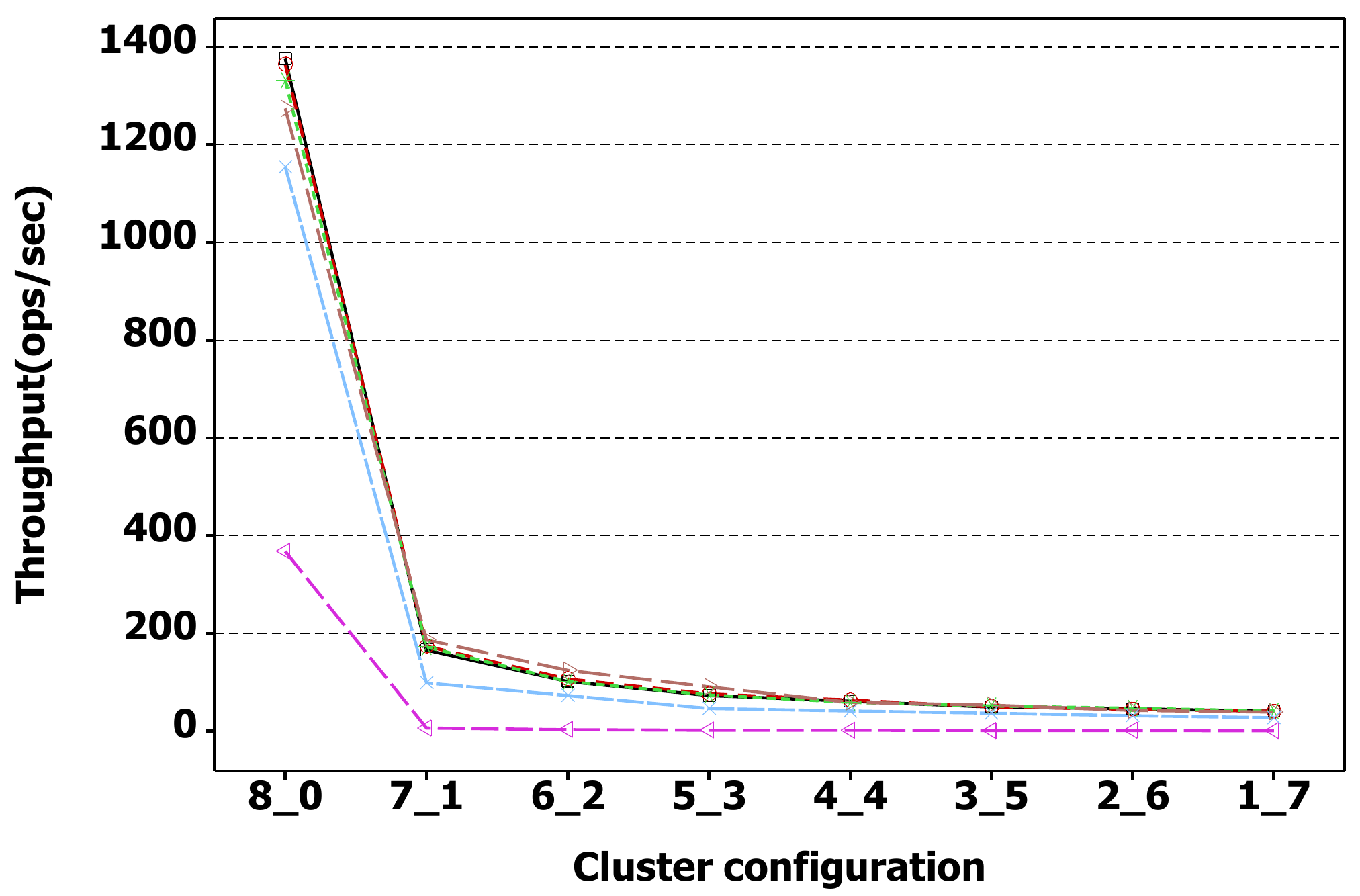}}
  \subfloat[MySQL]{\label{figur:mysql-thr}\includegraphics[width=0.5\textwidth]{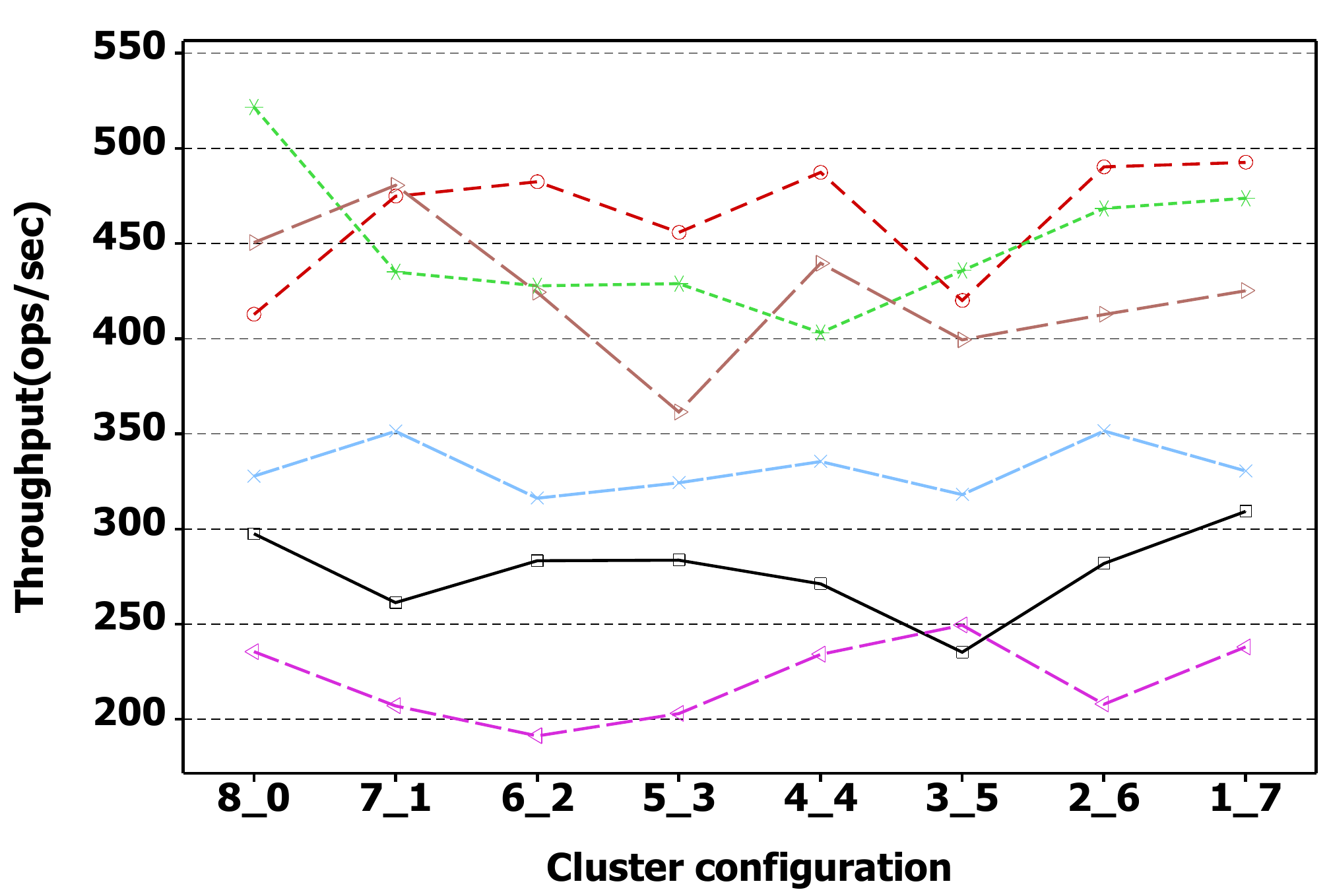}}
  \caption{Throughput for distributed databases. Value $n\_m$  in axis X represents $n$ nodes in the private cloud and $m$ nodes in the public cloud are deployed.}
\label{fig:throughput}
\end{figure*}

Fig. \ref{fig:throughput} illustrates the throughput of the six distributed databases against cluster configuration labelled with pairs of $(n\_m)$ in which $n$ is the number of VM instances exploited in the local OpenStack and $m$ is the number of VM instances burst into Azure cloud datacenter. For each database and cluster configuration, we used a freshly installed and established database cluster  and loaded the data. We refer to cluster configuration with pairs $(8\_0)$ and $(1\_7)$ as \textit{non-bursting} and \textit{full-bursting} respectively. All pairs, except (8\_0), are referred to as hybrid cluster configurations. It should be noted that in the full-bursting setting, we still exploit one VM instance in OpenStack due to keeping the definition of hybrid cloud.     

As depicted in Fig. \ref{figur:mongo-thr}, in the experiment with the non-bursting setting, MongoDB achieves the highest throughput between 550-630 ops/sec for read-related ( i.e., read-only, read-latest and read-intensive) workloads. This value reduces for the write-intensive workload (500 ops/sec), followed by scan and read-modify-write (400 ops/sec) workloads. As the cluster configuration changes from non-bursting to bursting (i.e., cluster configurations of $(7\_1)$, $(6\_2)$,..., $(1\_7)$), the throughput of MongoDB slightly improves for read-related workloads, and increases by 10\% for the read-modify-write workload. By contrary, for write-intensive and scan workloads, the throughput of MongoDB significantly reduces when more than half of the resources are provided through the public cloud. In precise, this value declines from 500 to 350 ops/sec for write-intensive and from 350 to 230  ops/sec for scan when cluster configuration changes from $(4\_4)$ to $(3\_5)$.  

As illustrated in Fig. \ref{figur:cassandra-thr},  Cassandra reaches 180-230 ops/sec for write- and read-latest workloads in non-bursting. For other workloads, these values decreases to 25-55 ops/sec. When Cassandra bursts into the public cloud, the throughput for write-intensive and read-latest workloads declines significantly, while its effect on  other workloads is less. With the increment in the  number of VM instances bursting into the public cloud, Cassandra's throughput diminishes in cluster configurations of $(5\_3)$ and $(2\_6)$ and increases in cluster configurations of $(6\_2)$ and $(3\_5)$. 

Figs. \ref{figur:riak-thr} and \ref{figur:couchdb-thr} represent the throughput for Riak and Couchdb databases. For non-bursting, Riak has a throughput of 500-750 ops/sec, while Couchdb has a throughput of 280-350 ops/sec for read-related workloads, and 100-140 ops/sec for the remaining workloads. Cloud bursting  leads to 50-100 ops/sec and less 50 ops/sec for Riak and Couchdb respectively. As can be seen, this performance is roughly constant for all workloads with different hybrid cluster configurations.  

Fig. \ref{figur:redis-thr} depicts the throughput for Redis databases. In non-bursting, Redis has the best performance in comparison to all databases due to being memory-based. Thus,  Redis is not optimized to work well under network-intensive workloads. The results show that cloud bursting is still effective if Redis is provisioned with public cloud for less than half of its required resources.

The throughput of MySQL is shown in Fig. \ref{figur:mysql-thr}. In the non-bursting setting, MySQL attains the best throughput for read-related workloads (420-530 ops/sec)  and the worst one for scan (280 ops/sec) and write-intensive (300 ops/sec) workloads. Similar to MongoDB,  MySQL's throughput increases for read-intensive and read-modify-write workloads as cluster configuration changes from non-bursting to bursting (i.e., different hybrid cluster configurations). Like MongoDB,  the throughput of MySQL for read-related workloads is better than for other workloads when cluster configuration varies from $(7\_1)$ to $(1\_7)$.  

\textbf{Discussion:} From the results discussed above, we make the following remarks. Under the conditions of the conducted experiments, MongoDB exhibits the best performance in throughput for different workloads for all hybrid cluster configurations. This performance is closely followed by MySQL with a reduction of 16\%-28\% for read-related workloads and a decrease of 34\%-43\% for write-intensive, scan, and read-modify-write workloads as the cluster configutration changes from $8\_0$ to $4\_4$. The high latency degrades the performance of Cassandra, Riak, and Couchdb  due to using the quorum-based technique for read and write  commit operations. Redis has also the same performance degeneration because it is not  designed to be deployed across WAN.   

\begin{figure*}[h!]
  \centering
  \subfloat[Mongo]{\label{figur:mongo-rl}\includegraphics[width=0.5\textwidth]{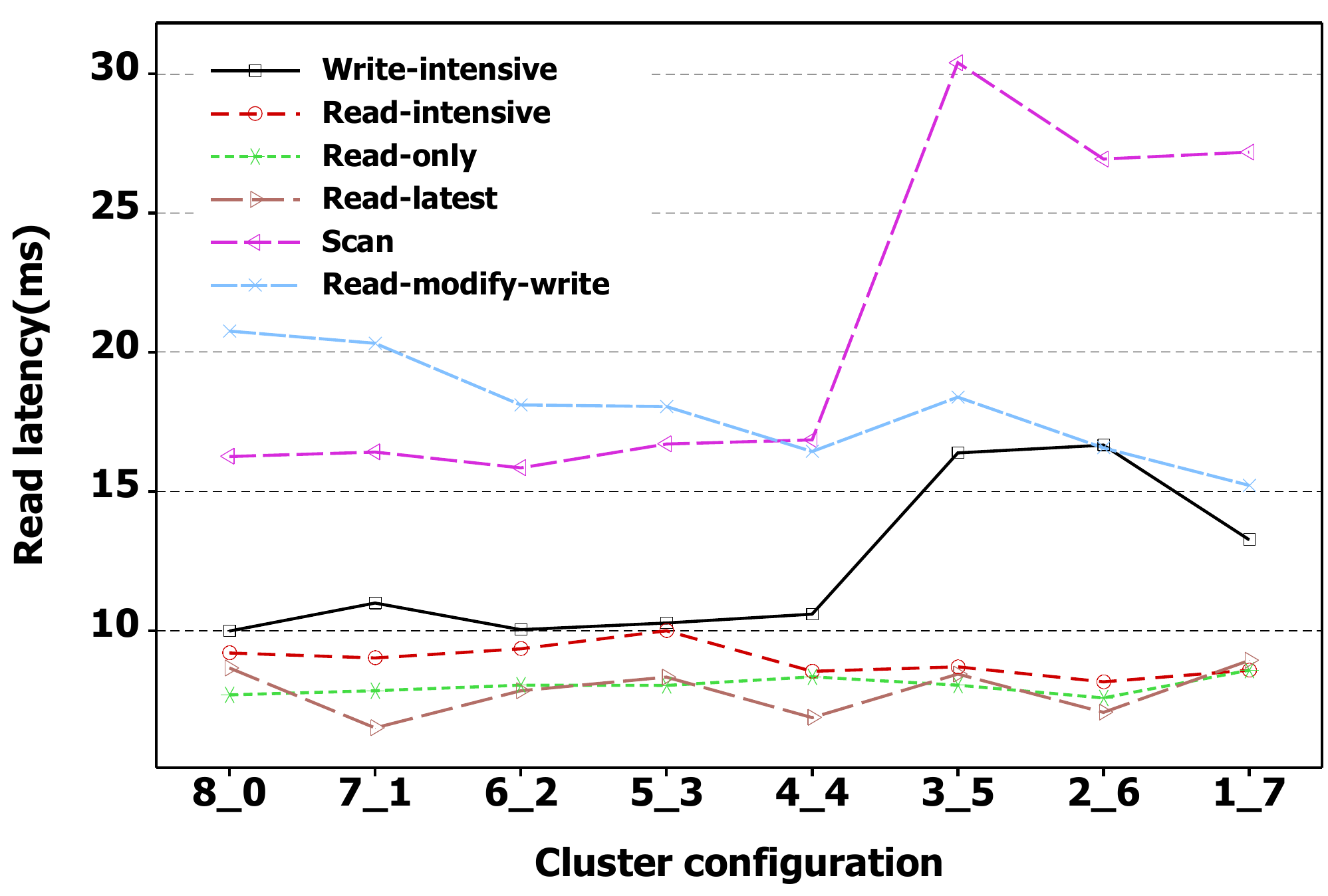}}
  \subfloat[Cassandra]{\label{figur:cassandra-rl}\includegraphics[width=0.5\textwidth]{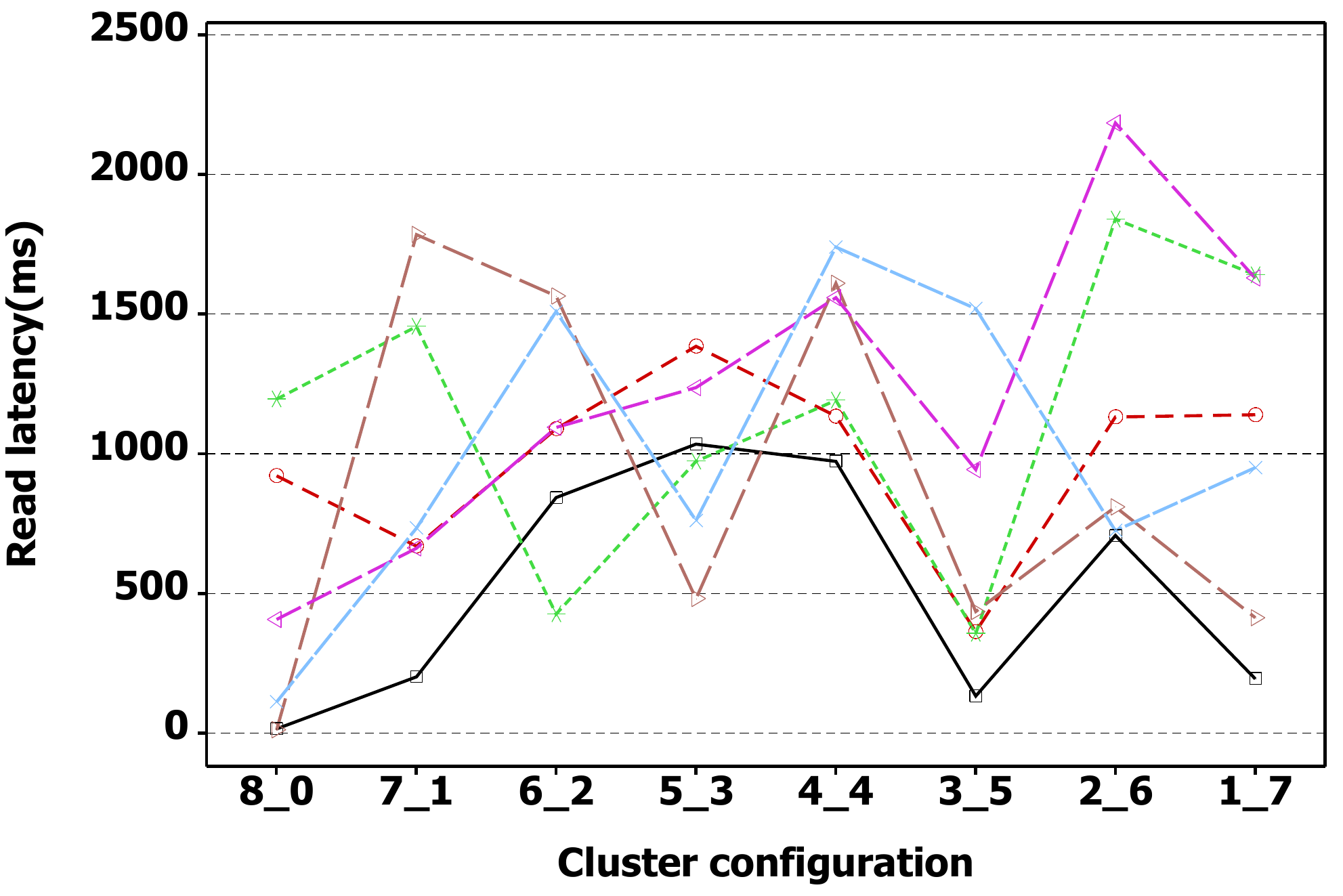}}\\
  \subfloat[Riak]{\label{figur:riak-rl}\includegraphics[width=0.5\textwidth]{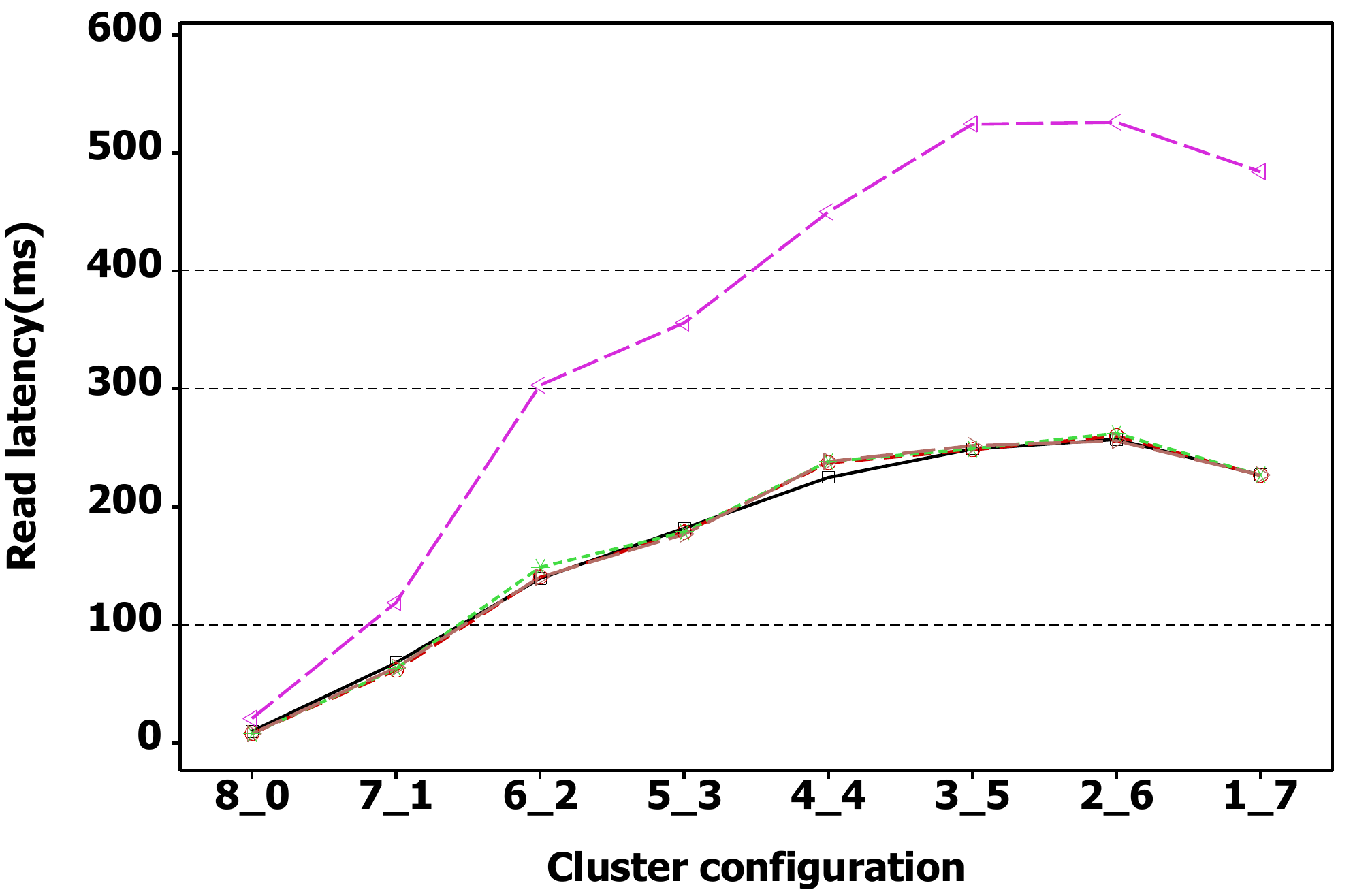}}
  \subfloat[Couchdb]{\label{figur:couchdb-rl}\includegraphics[width=0.5\textwidth]{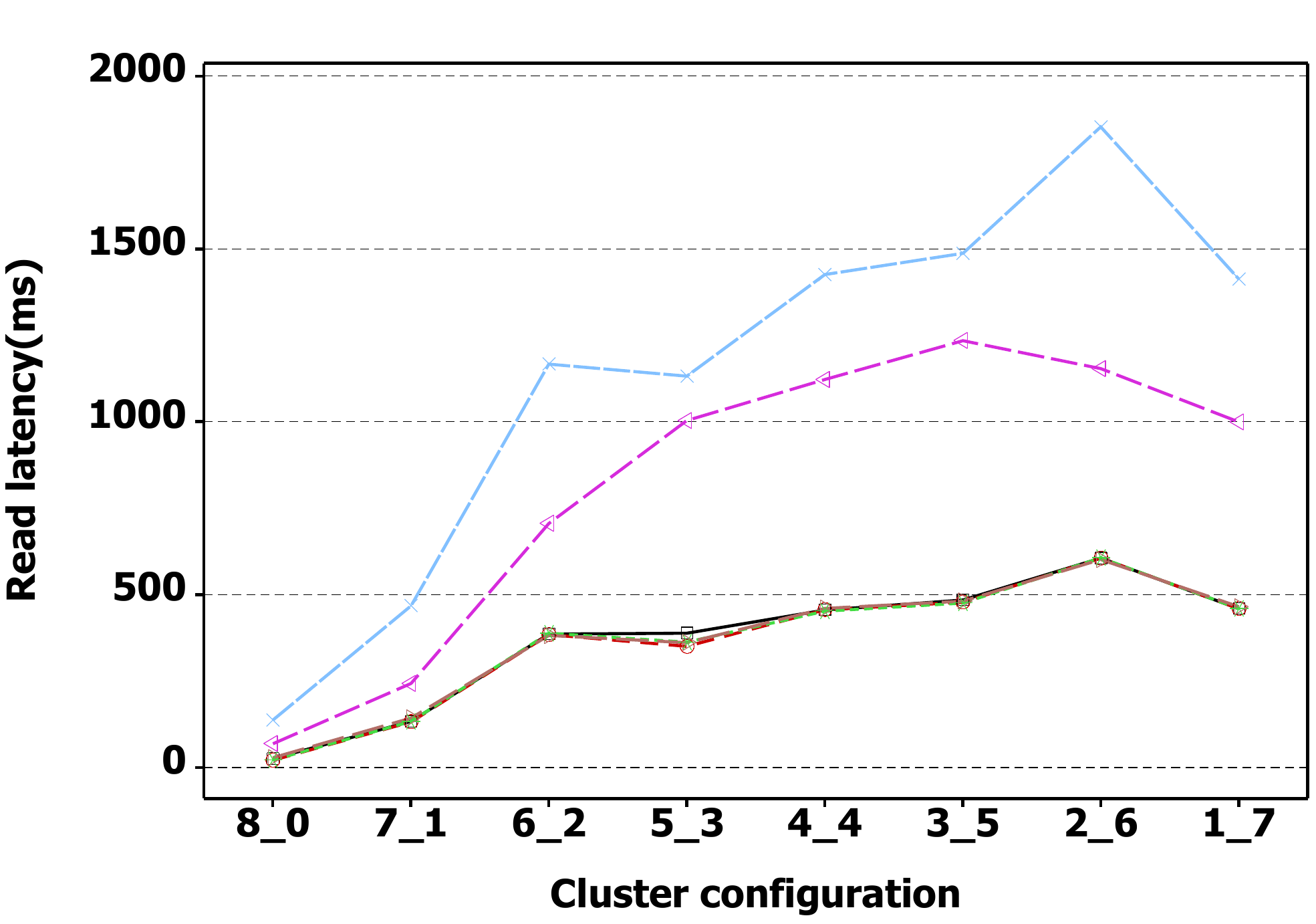}}\\
  \subfloat[Redis]{\label{figur:redis-rl}\includegraphics[width=0.5\textwidth]{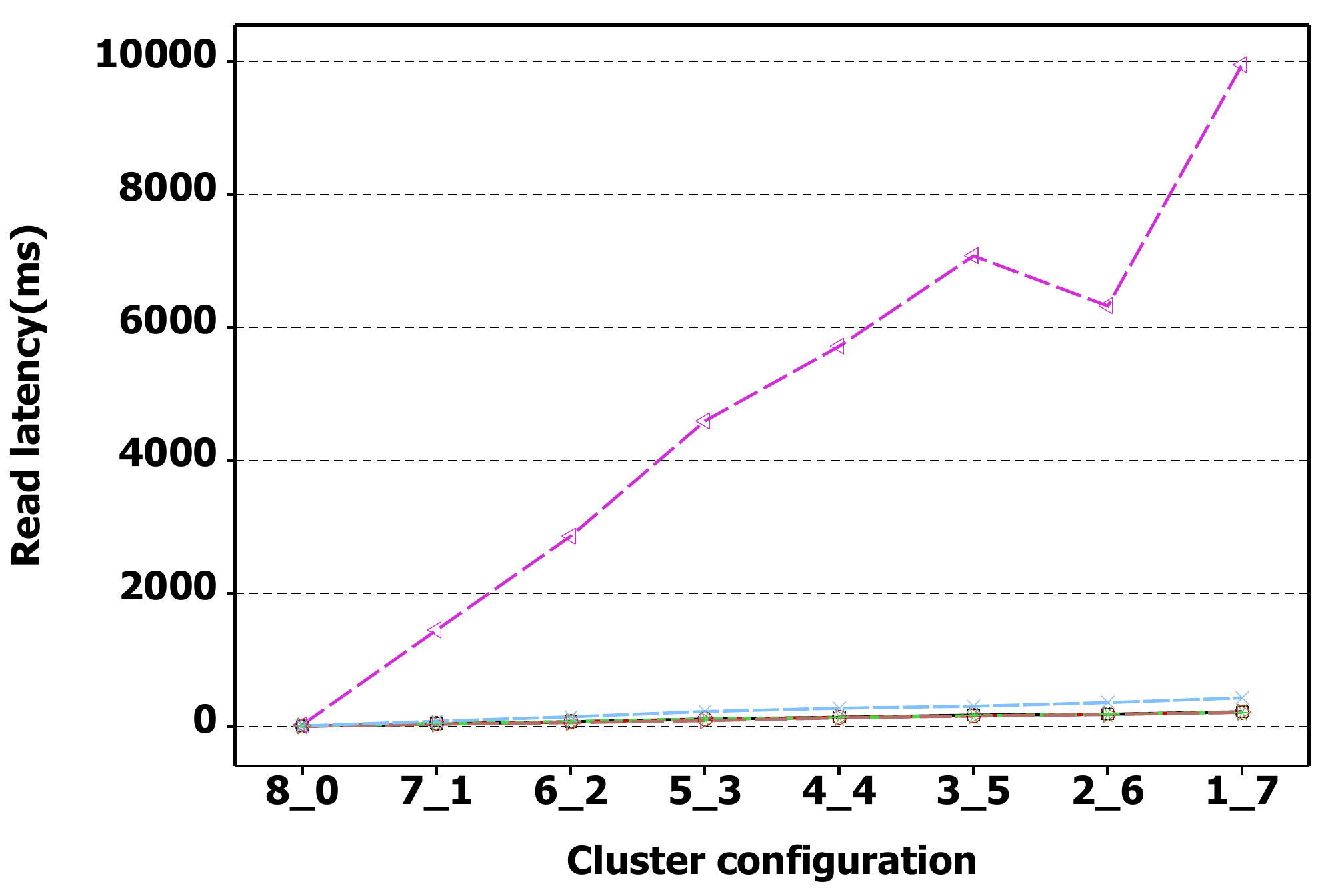}}
  \subfloat[MySQL]{\label{figur:mysql-rl}\includegraphics[width=0.5\textwidth]{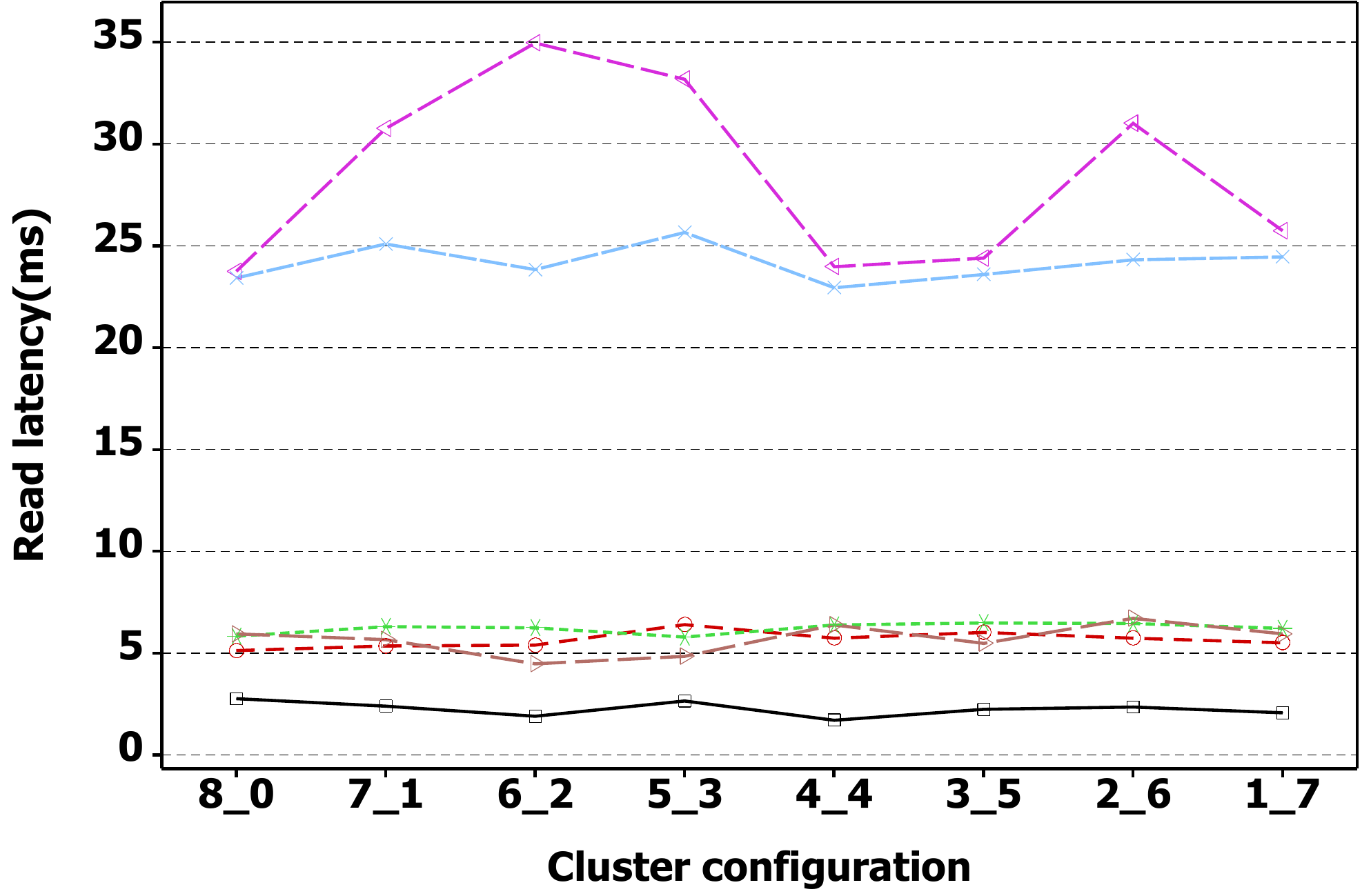}}
  \caption{Read latency for distributed databases. Values $n\_m$  in axis $X$ represent $n$ nodes in the private cloud and $m$ nodes in the public cloud.}
\label{fig:readlatency}
\end{figure*}

In Fig. \ref{fig:readlatency}, the read latency can be seen. The read latency in MongoDB with non-bursting is less than the one for the hybrid cluster configurations (i.e., $7\_1$ - $1\_7$) for all workloads except scan and write-intensive. The deployment of MongoDB with the hybrid cluster configurations, the read latency reduces by 25\% for the read-modify-write workload. By contrast, for write-intensive and scan, the read latency substantially increases as the cluster configuration changes from $(4\_4)$ to $(3\_5)$, followed by a reduction of 5\% for the cluster configuration of $(1\_7)$.

Fig. \ref{figur:cassandra-rl} illustrates the read latency of Cassandra. The read latency increases as the setting changes to the hybrid cluster configurations, which is consistent with its low throughput. This database has unstable behaviour in read latency due to reading data from remote datacenter based on the quorum-based technique.  One interesting observation of the results is that Cassandra achieves the lowest read latency in cluster configurations of $(5\_3)$ and $(3\_5)$ among all the hybrid cluster configurations. This might be because Cassandra makes a read commitment based on the quorum-based technique from three replicas as  default setting.  

%This might be due to the fact that quorum-based voting to read data from remote cloud impose high latency. 

As shown in Fig. \ref{figur:riak-rl}, Riak reduces the read latency  more than three times in comparison to the Cassandra for the cluster configurations of $(7\_1)$-$(2\_6)$. Couchdb is comparable in the read latency with Cassandra although it is more stable due to exploiting the \textit{local} quorum-based technique to commit operations (\ref{figur:couchdb-rl}). 
Fig. \ref{figur:redis-rl} illustrates the read latency for Redis. As expected, the Redis's read latency for all workloads (except scan) is the lowest in comparison to ones for other databases because Redis is an in-memory database. However, this database yields high read latency for scan workload as more nodes are exploited in the public cloud.

Like MongoDB, MySQL achieves low latency for read operations: about 2.5ms for write-intensive workload and 5ms  for  read-related workloads. These values roughly remain constant for all  hybrid cluster configurations (Fig. \ref{figur:mysql-rl}). This is because MySQL performs well in pairing replica nodes across clouds. However, MySQL imposes higher latency for scan (22ms-25ms) and read-modify-write (22ms-35ms). It should be noted that the read latency for scan workload reduces when the number of nodes, bursting into the public cloud, is a multiple of 2.

\begin{figure*}[h!]
  \centering
  \subfloat[Mongo]{\label{figur:mongo-wl}\includegraphics[width=0.5\textwidth]{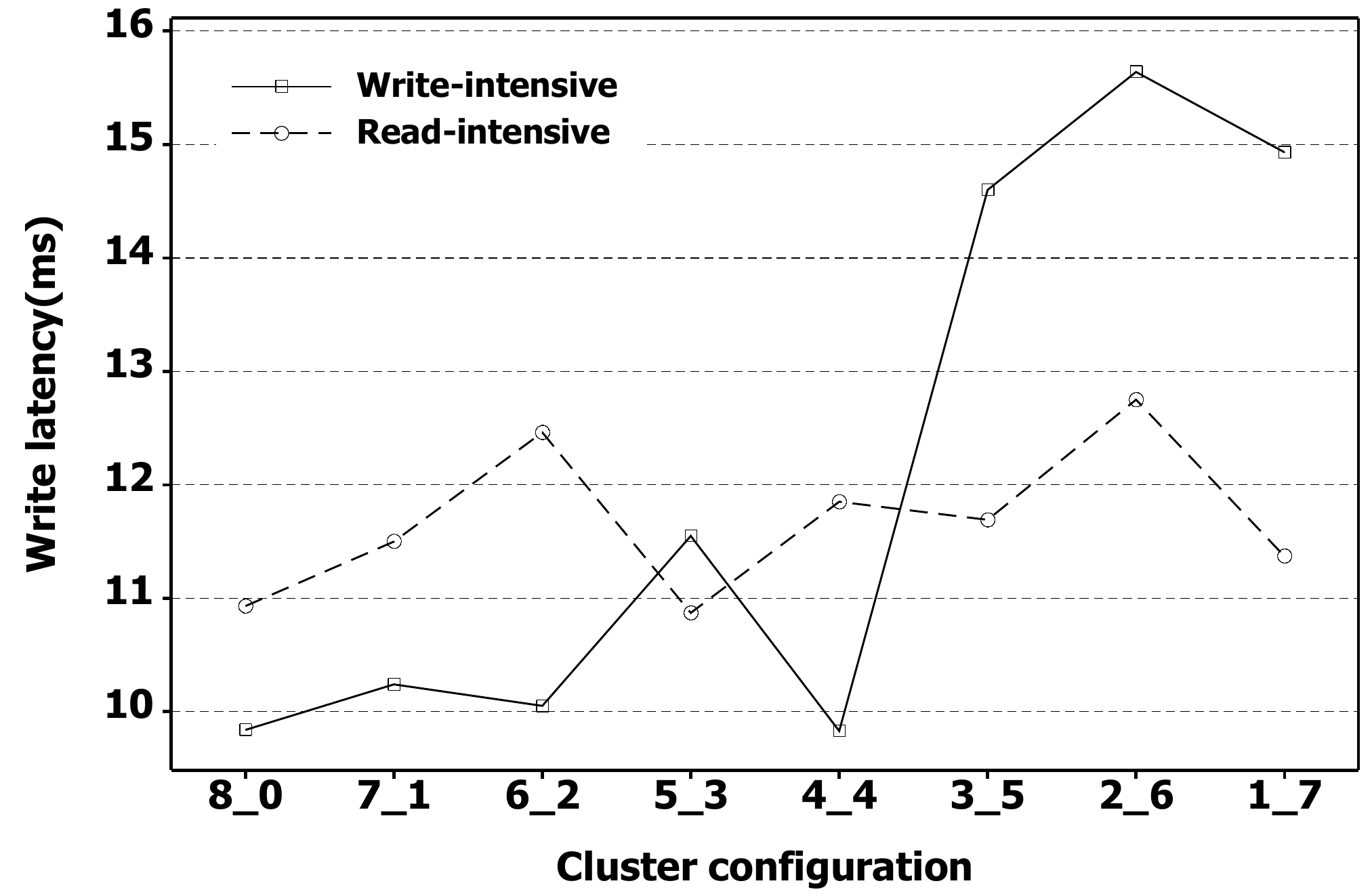}}
  \subfloat[Cassandra]{\label{figur:cassandra-wl}\includegraphics[width=0.5\textwidth]{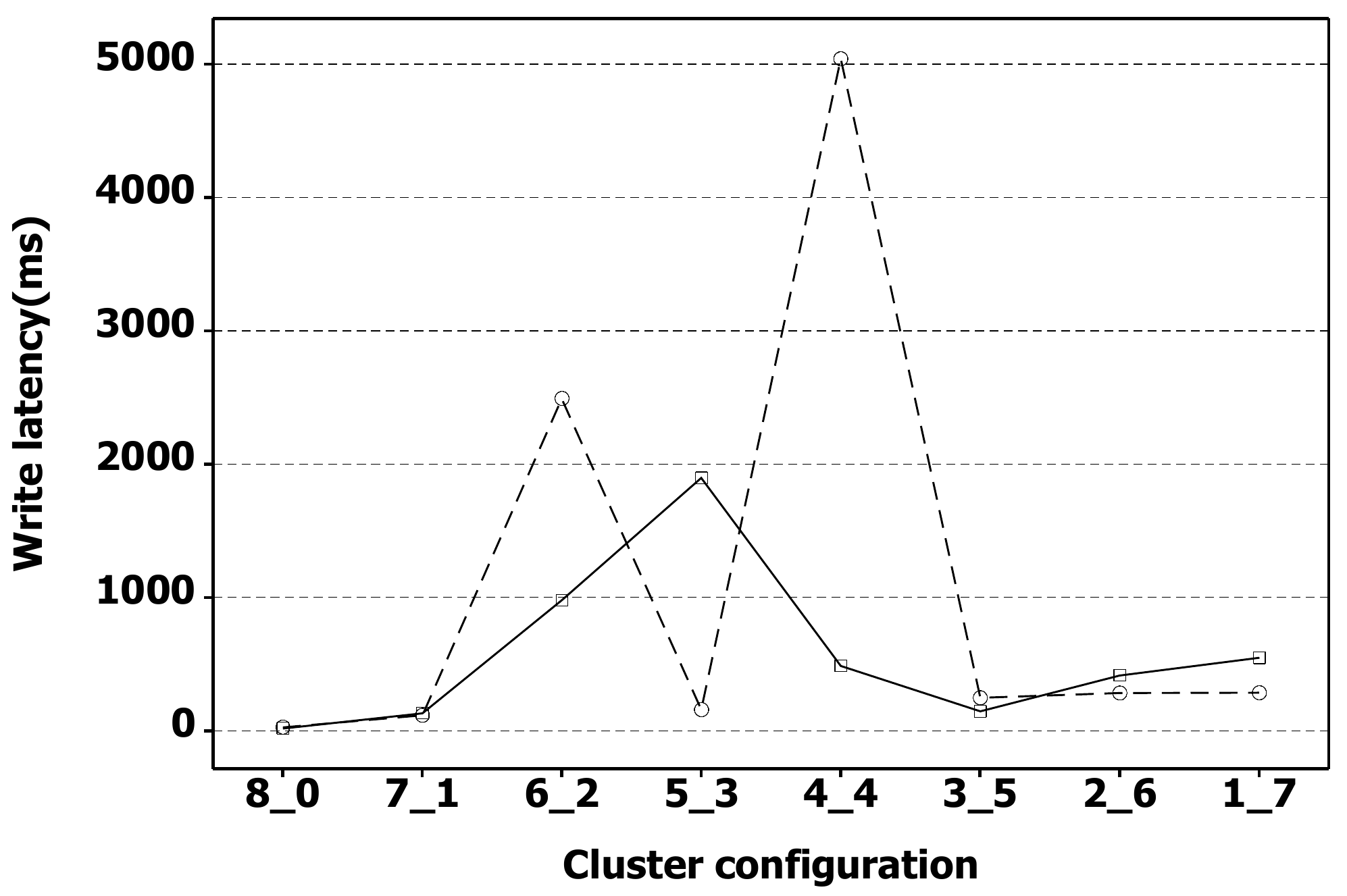}}\\
  \subfloat[Riak]{\label{figur:riak-wl}\includegraphics[width=0.5\textwidth]{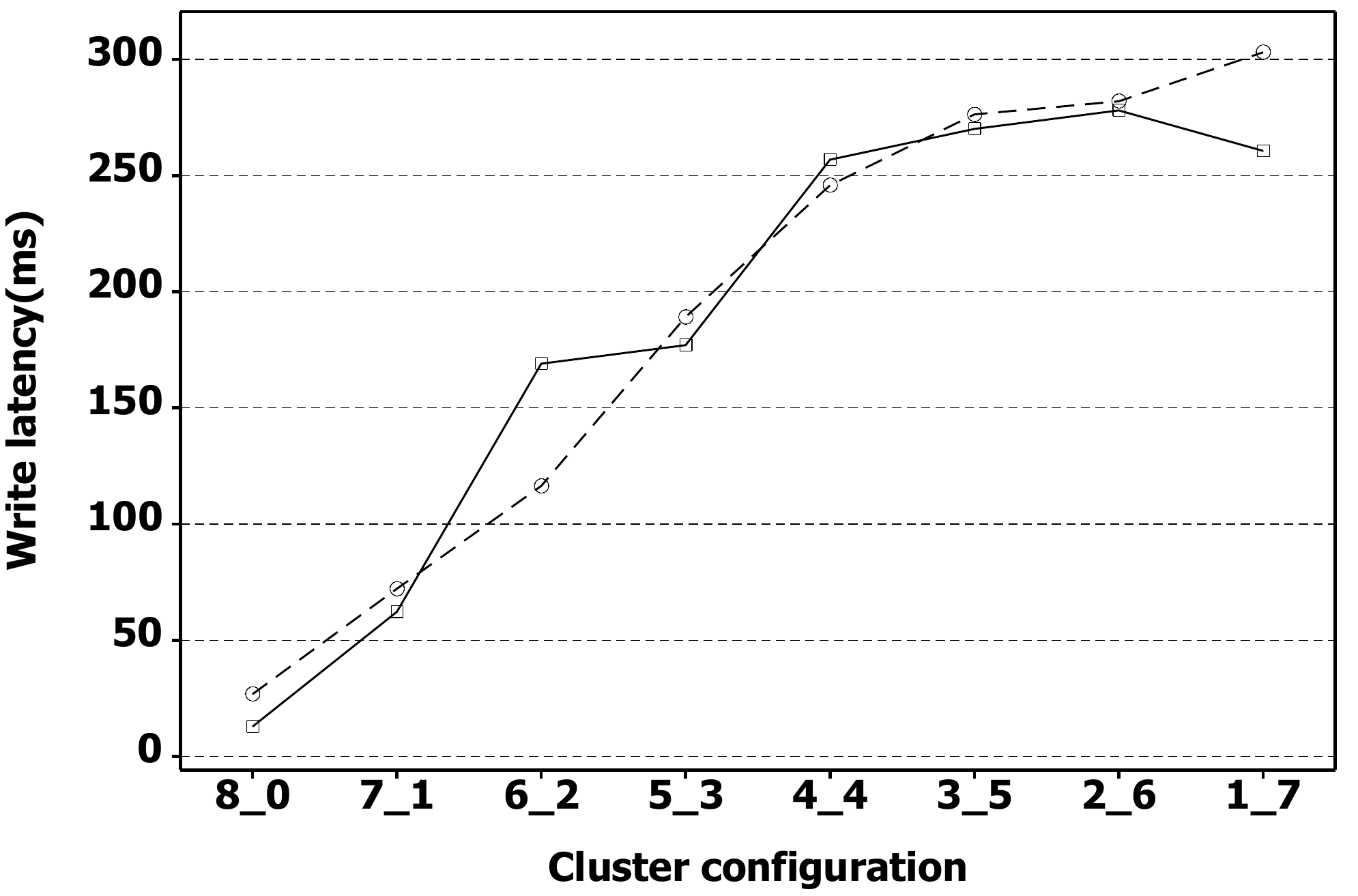}}
  \subfloat[Couchdb]{\label{figur:couchdb-wl}\includegraphics[width=0.5\textwidth]{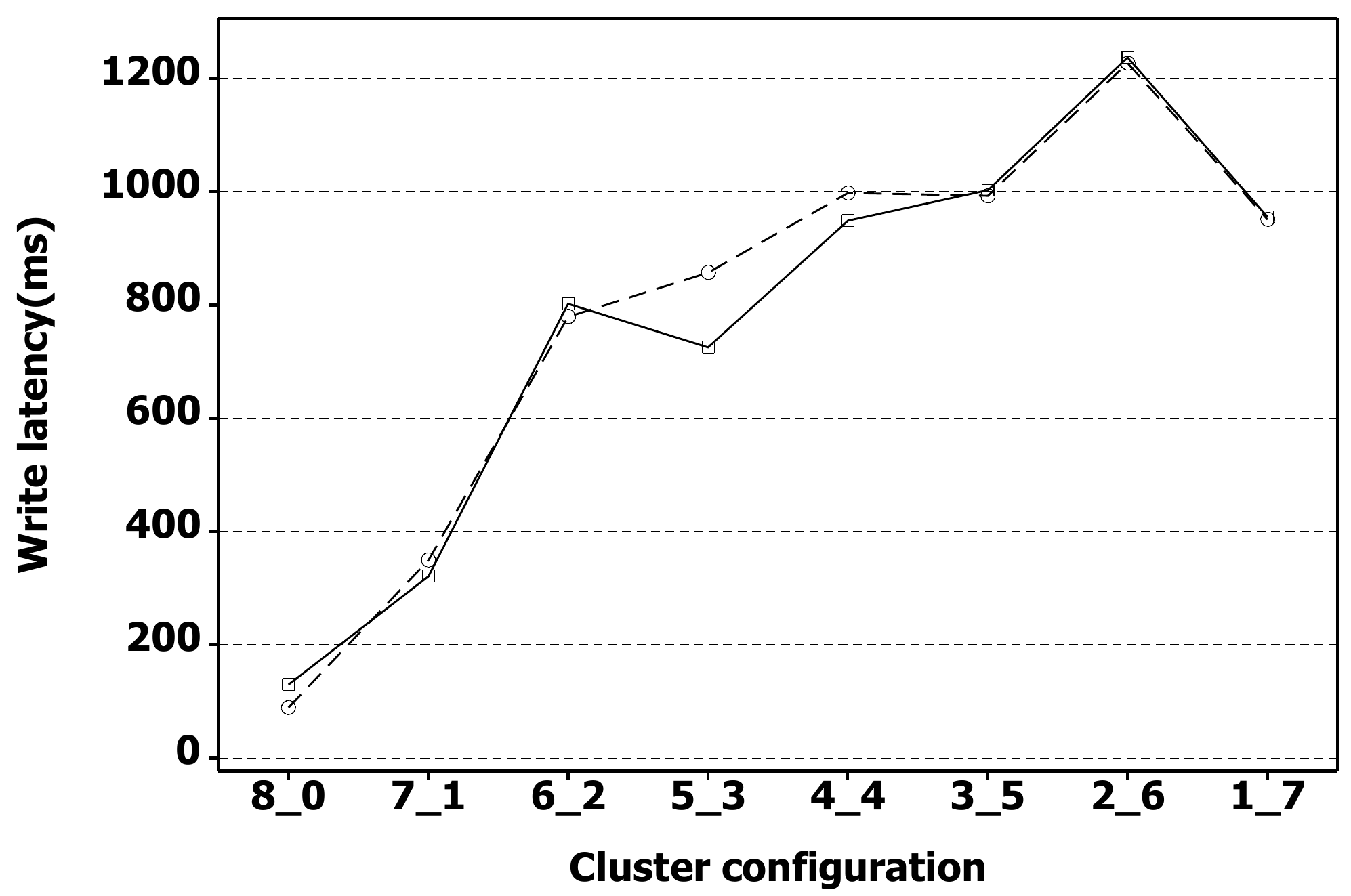}}\\
  \subfloat[Redis]{\label{figur:redis-wl}\includegraphics[width=0.5\textwidth]{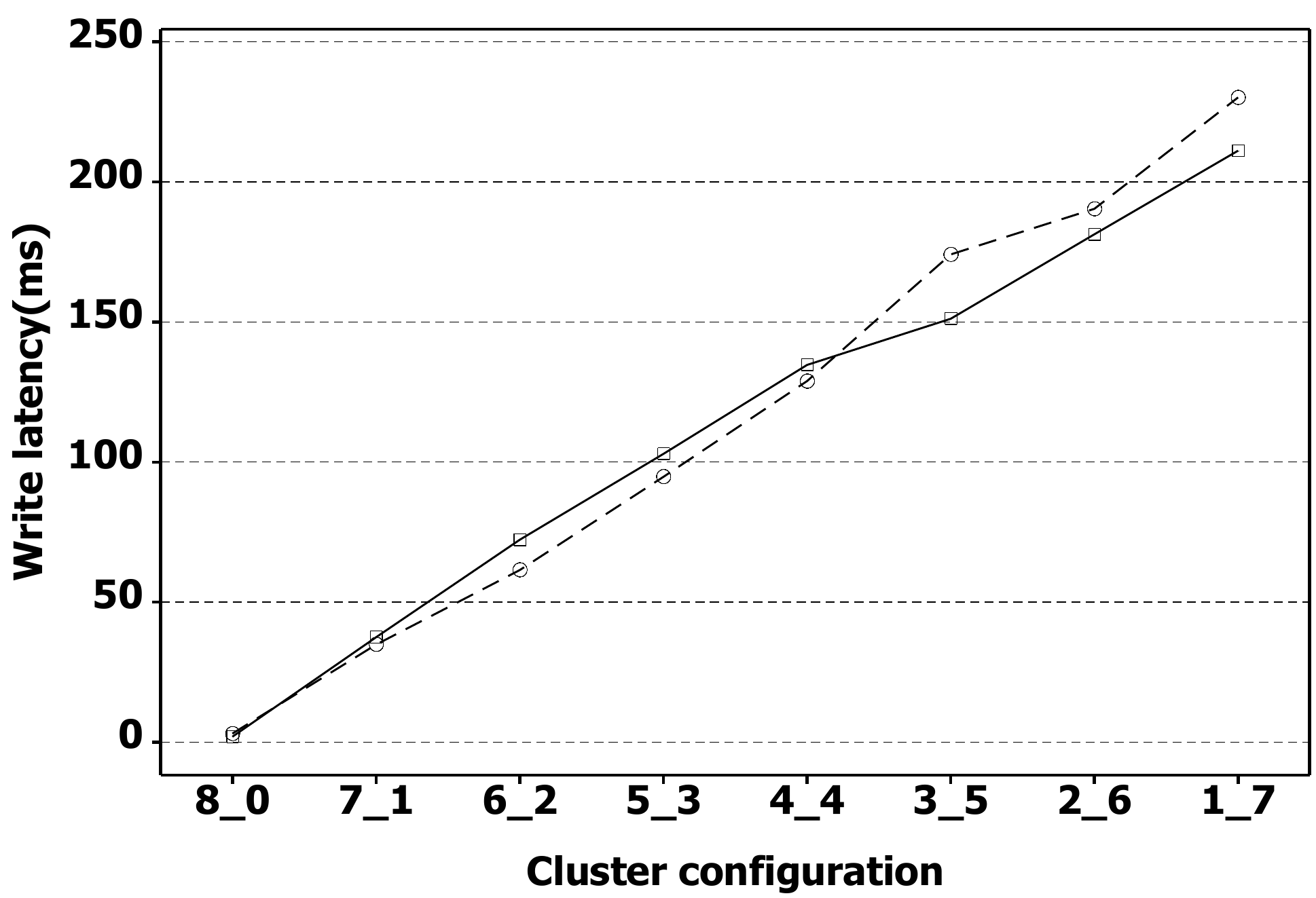}}
  \subfloat[MySQL]{\label{figur:mysql-wl}\includegraphics[width=0.5\textwidth]{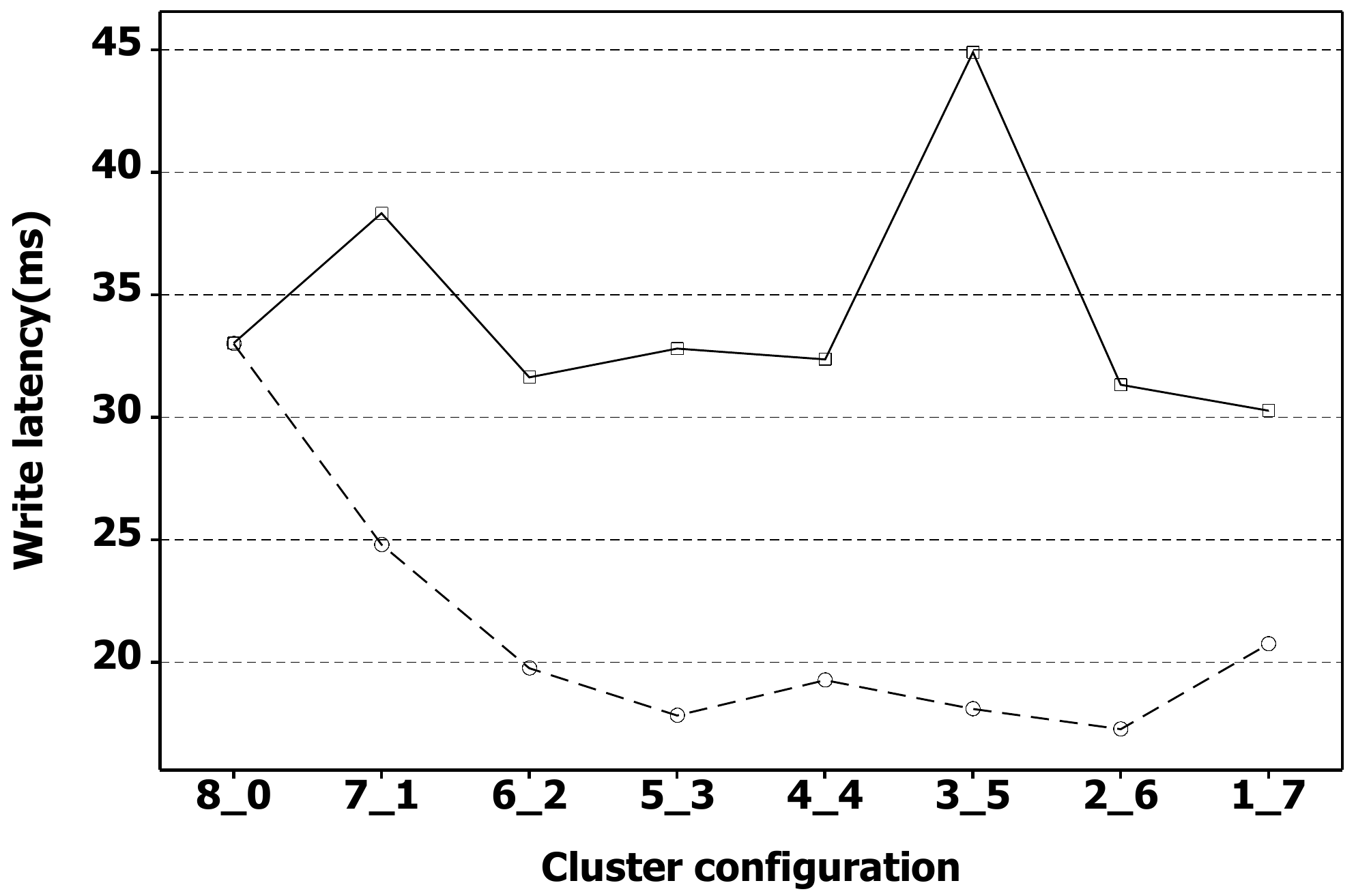}}
  \caption{Write latency for distributed databases. Values $n\_m$  in axis $X$ represent $n$ nodes in the private cloud and $m$ nodes in the public cloud.}
\label{fig:writelatency}
\end{figure*}

Fig. \ref{fig:writelatency} shows the write latency for the write- and read-intensive workloads including 50\% and 5\% write operations respectively. Among NoSQL databases, MongoDB  has the lowest write latency varying from 10ms-15ms. In this performance metric, Redis comes in the second rank  and its write latency increase as the number of Azure VM instances involved in the hybrid cloud raises. Riak  has lower write latency compared to Cassandra and Couchdb especially for the hybrid cluster configurations of $(7\_1)$-$(2\_6)$. In contrast to all databases, MySQL's write latency decreases as the cluster configuration changes from non-bursting to bursting. 

\begin{table}[t]
	% \begin{threeparttable}
	%\begin{adjustbox}{width=\textwidth}
	\caption{Percentage of error for operations operated on distributed databases}\label{tab:errorope}
	\centering
	%\vspace{-3mm}
	\begin{tabular}{p{3cm}p{3cm}p{3cm}p{4.2cm}}
		%\centering
		\hline
		Database name  &  Write-intensive &Read-intensive &Read-modify-write   \\\hline\hline
		MongoDB                 &11.33\%	&11.13\% &11.46\% \\
		Cassandra             &20\%	 &25\%	&26\%   \\
		Couchdb               &3.5\%	&0.27\%	&0\%  \\
		Redis               &0.27\%	&0.63\%	&0.77\%  \\
		Riak, MySQL    &0\%  &0\% &0\%    \\\hline
		\end{tabular}
	%\end{threeparttable}
	\vspace{-5mm}
\end{table}

\textbf{Discussion:} The overall results show that MongoDB and MySQL  have the lowest read latency, while Cassandra behaves in unstable way for all workloads. In respect to the write latency, MongoDB and MySQL again produce the lowest write latency although in some hybrid cluster configurations they have spikes. These spikes might be resulted from the network latency fluctuations, as shown in Fig.\ref{figur:op-az-across}.

Table \ref{tab:errorope} summarize the error percentage for operations performed on distributed databases in the case of hybrid cluster configurations. Error for one operation means that for some reasons (e.g., network partitioning or consistency violation) the YCSB client is not able to perform or verify an operation. Results show that Riak and MySQL are error-free in operations conducted by the YCSB client.  In contrast, Cassandra has the highest error percentage for operations (20\%-26\%), followed by MongoDB. These results are consistent with attaining the highest throughput by MongoDB deployment in the hybrid cloud. Surprisingly, although Cassandra has the highest percentage in error for operations, it could not outperform MongoDB in throughput. This might  because MongoDB supports eventual consistency while Cassandra offers quorum-based consistency by default. It is worth nothing that in the non-bursting setting, all databases  \ref{tab:errorope} are operating error-free.

\subsection{Practical observations}

We report additional observations and findings from a practitioner's perspective during the implementation of hybrid cloud and deployment of the six distributed databases. 

To make a secure connection between the local OpenStack and Microsoft Azure clouds, we tested both default Microsoft Azure VPN and WireGurd. The deployment of WireGuard is more simple because we only need to exchange private and public keys of the VMs involved. While in its counterpart, we need a public IP address in the Microsoft Azure for VPN gateway and require to configure many settings in the OpenStack side to establish connection. In addition, WireGuard is free and provides a reliable connection. In contrast, Microsoft Azure is costly and depends on the amounts of  data transferred. Another advantage of WireGuard is the high flexiability and universal applicability for other cloud providers such as AWS and Google. In short, due to these features, we selected WireGuard to make a secure connection. 

There are several ways to deploy virtualized cloud infrastructure. As an example,  Microsoft Azure  supports a number of tools such as Azure command-line interface (CLI), Azure Power Shell , and Azure Portal\footnote{Microsoft Azure: https://azure.microsoft.com/en-au/overview/}. However, these tools are appropriate when a limited number of resources need to be managed. In efforts to improve automation of virtulaized resource management several solutions such as Terraform and Ansible have been developed.  We selected Terraform due to high adoption, syntax simplicity, and multi-cloud support. Terraform enabled strongly reproducible  experimentation with high performance. Interestingly, infrastructure deployment in the local OpenStack cloud is significantly faster (up to a factor of 10) compared to Azure deployment.      

In respect to the deployment of 6 databases, we found that their installation and cluster configuration is different especially the configuration of NoSQL databases and MySQL. MongoDB and Cassandra databases follow a master-slave architecture for cluster configuration, while Redis, Riak and CouchDB treat all deployed nodes in the same way. In addition, some databases have further specific restrictions such as number of nodes for MySQL or number of master nodes for Cassandra. Similarly, authentication setup  required for  CouchDB and MySQL complicates unattended installation. These database-specific differences led to the necessity to design flexible and extendable installation scripts architecture. As a result, the developed deployment system is easily expandable to support other new databases.

Lastly, the benchmarking phase was complete using YCSB. Due to individual database differences, a set of database-specific YCSB invocation scripts have been developed. These included variation on authentication details, uri format, initialization queries, port numbers and so on.  A notable problem was a lack of compatibility with CouchDB. Thus, patching and recompiling the CouchDB YCSB connector was required\footnote{YCSB-CouchDB-Binding: https://github.com/akhildixit/YCSB-CouchDB-Binding}.  Additional saturation-related consideration raised by \cite{Rabl2012} was not applicable for our experiment scenarios due to a limited bandwidth and high latency between the public and private clouds.

\section{Conclusions and Future Work}\label{sec:conclusion}
In this paper, we presented an automation of hybrid cloud implementation using (i) WireGuard as a Linux-based VPN to make a secure connection between public and private clouds, and (ii) Terraform as a software tool to deploy infrastructure resources based on the required number of VM instances, VM flavour, security group network, sub-network, etc. We deployed the implemented hybrid cloud to evaluate cloud-bursting at the level of distributed databases. Our evaluation reveals that MongoDB and MySQL Cluster work well in throughput and latency of read and write operation as if they burst into the public cloud. In contrast, Cassandra, Riak, Couchdb, and Redis, exhibit low performance especially when they supply more than half of their resources via cloud bursting across clouds locating at a long distance from each other.  This leads to high and unpredictable latency that negatively affects the performance of these databases. Thus, it is required either to deploy such databases across data centers locating at a short distance (i..e, within a continent), or improve them in data model, data sharding and replication policy. Overall, while experiments  revealed that cloud-bursting does not improve the database performance, under certain condition capacity improvement might still be beneficial. 

For future work, we plan to investigate the static and on-the-fly dynamic up and down scaling  of distributed databases deployment in hybrid clouds. Additionally, we are interested in determining the impact of distance between the private and public clouds and replication factor on the throughput and latency of operations in our study. Finally, we plan to extend our tested architecture across different regions of the public  cloud deployment and design optimally data placement algorithms to improve performance with the fixed and variable number of VM instances.  

\section*{Acknowledgment}
We thank Faheem Ullah and anonymous reviewers for their valuable comments in improving the quality of the paper.

\bibliographystyle{elsarticle-num}
\bibliography{references}

\begin{thebibliography}{26}
\providecommand{\natexlab}[1]{#1}
\providecommand{\url}[1]{\texttt{#1}}
\expandafter\ifx\csname urlstyle\endcsname\relax
  \providecommand{\doi}[1]{doi: #1}\else
  \providecommand{\doi}{doi: \begingroup \urlstyle{rm}\Url}\fi

\bibitem[Abdi et~al.(2017)Abdi, PourKarimi, Ahmadi, and Zargari]{Abdi2017}
S.~Abdi, L.~PourKarimi, M.~Ahmadi, and F.~Zargari.
\newblock Cost minimization for deadline-constrained bag-of-tasks applications
  in federated hybrid clouds.
\newblock \emph{Future Gener. Comput. Syst.}, 71\penalty0 (C):\penalty0
  113--128, June 2017.
\newblock ISSN 0167-739X.

\bibitem[Abramova and Bernardino(2013)]{Abramova2013}
V.~Abramova and J.~Bernardino.
\newblock Nosql databases: Mongodb vs cassandra.
\newblock In \emph{Proceedings of the International C* Conference on Computer
  Science and Software Engineering}, C3S2E '13, pages 14--22, New York, NY,
  USA, 2013. ACM.
\newblock ISBN 978-1-4503-1976-8.

\bibitem[Armbrust et~al.(2010)Armbrust, Fox, Griffith, Joseph, Katz, Konwinski,
  Lee, Patterson, Rabkin, Stoica, and Zaharia]{armbrust2010}
M.~Armbrust, A.~Fox, R.~Griffith, A.~Joseph, R.~Katz, A.~Konwinski, G.~Lee,
  D.~Patterson, A.~Rabkin, I.~Stoica, and M.~Zaharia.
\newblock A view of cloud computing.
\newblock \emph{Commun. ACM}, 53:\penalty0 50--58, 04 2010.
\newblock \doi{10.1145/1721654.1721672}.

\bibitem[Buyya et~al.(2009)Buyya, Yeo, Venugopal, Broberg, and
  Brandic]{buyya2009}
R.~Buyya, C.~S. Yeo, S.~Venugopal, J.~Broberg, and I.~Brandic.
\newblock Cloud computing and emerging it platforms: Vision, hype, and reality
  for delivering computing as the 5th utility.
\newblock \emph{Future Generation Computer Systems}, 25\penalty0 (6):\penalty0
  599 -- 616, 2009.
\newblock ISSN 0167-739X.

\bibitem[Calheiros et~al.(2011)Calheiros, Ranjan, Beloglazov, De~Rose, and
  Buyya]{Calheiros2011}
R.~N. Calheiros, R.~Ranjan, A.~Beloglazov, C.~A.~F. De~Rose, and R.~Buyya.
\newblock Cloudsim: A toolkit for modeling and simulation of cloud computing
  environments and evaluation of resource provisioning algorithms.
\newblock \emph{Softw. Pract. Exper.}, 41\penalty0 (1):\penalty0 23--50, Jan.
  2011.
\newblock ISSN 0038-0644.
\newblock \doi{10.1002/spe.995}.

\bibitem[Calheiros et~al.(2012)Calheiros, Vecchiola, Karunamoorthy, and
  Buyya]{CALHEIROS2012}
R.~N. Calheiros, C.~Vecchiola, D.~Karunamoorthy, and R.~Buyya.
\newblock The aneka platform and qos-driven resource provisioning for elastic
  applications on hybrid clouds.
\newblock \emph{Future Generation Computer Systems}, 28\penalty0 (6):\penalty0
  861 -- 870, 2012.
\newblock ISSN 0167-739X.
\newblock Including Special sections SS: Volunteer Computing and Desktop Grids
  and SS: Mobile Ubiquitous Computing.

\bibitem[Cooper et~al.(2010)Cooper, Silberstein, Tam, Ramakrishnan, and
  Sears]{Cooper2010}
B.~F. Cooper, A.~Silberstein, E.~Tam, R.~Ramakrishnan, and R.~Sears.
\newblock Benchmarking cloud serving systems with ycsb.
\newblock In \emph{Proceedings of the 1st ACM Symposium on Cloud Computing},
  SoCC '10, pages 143--154, New York, NY, USA, 2010. ACM.
\newblock ISBN 978-1-4503-0036-0.

\bibitem[{Dhall} et~al.(2012){Dhall}, {Dhall}, {Batra}, and {Rani}]{Dhall2012}
H.~{Dhall}, D.~{Dhall}, S.~{Batra}, and P.~{Rani}.
\newblock Implementation of ipsec protocol.
\newblock In \emph{2012 Second International Conference on Advanced Computing
  Communication Technologies}, pages 176--181, Jan 2012.

\bibitem[{Donenfeld}(2018)]{Donenfeld2018}
J.~A. {Donenfeld}.
\newblock Wireguard: Next generation kernel network tunnel.
\newblock In \emph{white paper}, pages 1--20, Jun 2018.
\newblock \doi{https://www.wireguard.com/papers/wireguard.pdf}.

\bibitem[{Jing Han} et~al.(2011){Jing Han}, {Haihong E}, {Guan Le}, and {Jian
  Du}]{Han2011}
{Jing Han}, {Haihong E}, {Guan Le}, and {Jian Du}.
\newblock Survey on nosql database.
\newblock In \emph{2011 6th International Conference on Pervasive Computing and
  Applications}, pages 363--366, Oct 2011.

\bibitem[Klein et~al.(2015)Klein, Gorton, Ernst, Donohoe, Pham, and
  Matser]{Klein2015}
J.~Klein, I.~Gorton, N.~Ernst, P.~Donohoe, K.~Pham, and C.~Matser.
\newblock Performance evaluation of nosql databases: A case study.
\newblock In \emph{Proceedings of the 1st Workshop on Performance Analysis of
  Big Data Systems}, PABS '15, pages 5--10, New York, NY, USA, 2015. ACM.
\newblock ISBN 978-1-4503-3338-2.

\bibitem[Kuhlenkamp et~al.(2014)Kuhlenkamp, Klems, and
  R\"{o}ss]{Kuhlenkamp2014}
J.~Kuhlenkamp, M.~Klems, and O.~R\"{o}ss.
\newblock Benchmarking scalability and elasticity of distributed database
  systems.
\newblock \emph{Proc. VLDB Endow.}, 7\penalty0 (12):\penalty0 1219--1230, Aug.
  2014.
\newblock ISSN 2150-8097.

\bibitem[Li et~al.(2018)Li, Tang, and Luo]{Li2018}
C.~Li, J.~Tang, and Y.~Luo.
\newblock Towards operational cost minimization for cloud bursting with
  deadline constraints in hybrid clouds.
\newblock \emph{Cluster Computing}, 21\penalty0 (4):\penalty0 2013--2029, Dec
  2018.

\bibitem[{Li} and {Manoharan}(2013)]{Li2013}
Y.~{Li} and S.~{Manoharan}.
\newblock A performance comparison of sql and nosql databases.
\newblock In \emph{2013 IEEE Pacific Rim Conference on Communications,
  Computers and Signal Processing (PACRIM)}, pages 15--19, Aug 2013.
\newblock \doi{10.1109/PACRIM.2013.6625441}.

\bibitem[{Lima} et~al.(2016){Lima}, {Oliveira}, {Kieckbusch}, {Holanda},
  {Walter}, {Araújo}, {Victorino}, {Silva}, and {Lifschitz}]{Lima2016}
I.~{Lima}, M.~{Oliveira}, D.~{Kieckbusch}, M.~{Holanda}, M.~E. M.~T. {Walter},
  A.~{Araújo}, M.~{Victorino}, W.~M.~C. {Silva}, and S.~{Lifschitz}.
\newblock An evaluation of data replication for bioinformatics workflows on
  nosql systems.
\newblock In \emph{2016 IEEE International Conference on Bioinformatics and
  Biomedicine (BIBM)}, pages 896--901, Dec 2016.

\bibitem[{Loreti} and {Ciampolini}(2015)]{Loreto2015}
D.~{Loreti} and A.~{Ciampolini}.
\newblock A hybrid cloud infrastructure for big data applications.
\newblock In \emph{2015 IEEE 17th International Conference on High Performance
  Computing and Communications, 2015 IEEE 7th International Symposium on
  Cyberspace Safety and Security, and 2015 IEEE 12th International Conference
  on Embedded Software and Systems}, pages 1713--1718, Aug 2015.

\bibitem[Mansouri et~al.(2017)Mansouri, Toosi, and Buyya]{Mansouri2017a}
Y.~Mansouri, A.~N. Toosi, and R.~Buyya.
\newblock Data storage management in cloud environments: Taxonomy, survey, and
  future directions.
\newblock \emph{ACM Comput. Surv.}, 50\penalty0 (6):\penalty0 91:1--91:51, Dec.
  2017.
\newblock ISSN 0360-0300.
\newblock \doi{10.1145/3136623}.

\bibitem[Moschakis and Karatza(2015)]{Moschakis2015}
I.~A. Moschakis and H.~D. Karatza.
\newblock Multi-criteria scheduling of bag-of-tasks applications on
  heterogeneous interlinked clouds with simulated annealing.
\newblock \emph{J. Syst. Softw.}, 101\penalty0 (C):\penalty0 1--14, Mar. 2015.
\newblock ISSN 0164-1212.

\bibitem[Rabl et~al.(2012)Rabl, G\'{o}mez-Villamor, Sadoghi, Munt{\'e}s-Mulero,
  Jacobsen, and Mankovskii]{Rabl2012}
T.~Rabl, S.~G\'{o}mez-Villamor, M.~Sadoghi, V.~Munt{\'e}s-Mulero, H.-A.
  Jacobsen, and S.~Mankovskii.
\newblock Solving big data challenges for enterprise application performance
  management.
\newblock \emph{Proc. VLDB Endow.}, 5\penalty0 (12):\penalty0 1724--1735, Aug.
  2012.
\newblock ISSN 2150-8097.

\bibitem[{Rimal} et~al.(2009){Rimal}, {Choi}, and {Lumb}]{rimal2009}
B.~P. {Rimal}, E.~{Choi}, and I.~{Lumb}.
\newblock A taxonomy and survey of cloud computing systems.
\newblock In \emph{2009 Fifth International Joint Conference on INC, IMS and
  IDC}, pages 44--51, Aug 2009.

\bibitem[Toosi et~al.(2018)Toosi, Sinnott, and Buyya]{NADJARANTOOSI2018}
A.~N. Toosi, R.~Sinnott, and R.~Buyya.
\newblock Resource provisioning for data-intensive applications with deadline
  constraints on hybrid clouds using aneka.
\newblock \emph{Future Generation Computer Systems}, 79:\penalty0 765 -- 775,
  2018.
\newblock ISSN 0167-739X.

\bibitem[Tuli et~al.(2020)Tuli, Sandhu, and Buyya]{Tuli2020}
S.~Tuli, R.~Sandhu, and R.~Buyya.
\newblock Shared data-aware dynamic resource provisioning and task scheduling
  for data intensive applications on hybrid clouds using aneka.
\newblock \emph{Future Generation Computer Systems}, 106:\penalty0 595 -- 606,
  2020.
\newblock ISSN 0167-739X.

\bibitem[Vecchiola et~al.(2012)Vecchiola, Calheiros, Karunamoorthy, and
  Buyya]{VECCHIOLA2012}
C.~Vecchiola, R.~N. Calheiros, D.~Karunamoorthy, and R.~Buyya.
\newblock Deadline-driven provisioning of resources for scientific applications
  in hybrid clouds with aneka.
\newblock \emph{Future Generation Computer Systems}, 28\penalty0 (1):\penalty0
  58 -- 65, 2012.
\newblock ISSN 0167-739X.

\bibitem[Xiong et~al.(2016)Xiong, Yeliang, Lipeng, Bin, Song, and
  Dong]{XIONG2016}
F.~Xiong, C.~Yeliang, Z.~Lipeng, H.~Bin, D.~Song, and W.~Dong.
\newblock Deadline based scheduling for data-intensive applications in clouds.
\newblock \emph{The Journal of China Universities of Posts and
  Telecommunications}, 23\penalty0 (6):\penalty0 8 -- 15, 2016.
\newblock ISSN 1005-8885.

\bibitem[Xu et~al.(2017)Xu, Zhao, Ruan, Zhang, Tian, Dou, and Liu]{xu2017}
X.~Xu, X.~Zhao, F.~Ruan, J.~Zhang, W.~Tian, W.~Dou, and A.~Liu.
\newblock Data placement for privacy-aware applications over big data in hybrid
  clouds.
\newblock \emph{Security and Communication Networks}, 2017:\penalty0 1--15, 11
  2017.
\newblock \doi{10.1155/2017/2376484}.

\bibitem[Zhou et~al.(2019)Zhou, Wang, Cong, Lu, Wei, and Chen]{Zhou2019}
J.~Zhou, T.~Wang, P.~Cong, P.~Lu, T.~Wei, and M.~Chen.
\newblock Cost and makespan-aware workflow scheduling in hybrid clouds.
\newblock \emph{Journal of Systems Architecture}, 100:\penalty0 101631, 2019.
\newblock ISSN 1383-7621.

\end{thebibliography}

\end{document}